\documentclass[journal,twoside,web]{ieeecolor}
\usepackage{generic}
\usepackage{cite}
\usepackage{amsmath,amssymb,amsfonts}
\usepackage{algorithmic}
\usepackage{graphicx}
\usepackage{algorithm,algorithmic}
\usepackage{textcomp}
\usepackage{multirow} % Required for multirows
\usepackage{multicol} % Required for multirows
\usepackage{threeparttable} % 引入包
\usepackage[colorlinks=true, allcolors=blue]{hyperref} % 超链接
\usepackage{cleveref} % 智能引用
\usepackage{footnote}

\usepackage[switch]{lineno}

\usepackage[defaultcolor=red]{changes}

\usepackage{amssymb}
\usepackage{bbding}
\usepackage{booktabs}
\def\BibTeX{{\rm B\kern-.05em{\sc i\kern-.025em b}\kern-.08em
    T\kern-.1667em\lower.7ex\hbox{E}\kern-.125emX}}
\begin{document}

\title{HealthiVert-GAN: A Novel Framework of Pseudo-Healthy Vertebral Image Synthesis for Interpretable Compression Fracture Grading}
\author{Qi Zhang, Cheng Chuang, Shunan Zhang, Ziqi Zhao, Kun Wang, Jun Xu*, and Jianqi Sun*
\thanks{This research was supported by the National Natural Science Foundation of China (Grant No.62471297), the National Key R\&D Program of China (Grant No. 2021YFF0703702), the National Key Research and Development Program (Grant No. 2023YFC2411401), Shanghai Jiao Tong University Medical Engineering Cross Research Funds (Grant No. YG2025ZD03), and Shanghai Zhangjiang National Independent Innovation Demonstration Zone Special Development Fund Major Project (Grant No. ZJ2021-ZD-007). }
\thanks{Qi Zhang, Shunan Zhang, Ziqi Zhao, Jianqi Sun are with the School of Biomedical Engineering, Shanghai Jiao Tong University, Shanghai, China.}
\thanks{Cheng Chuang is with Kashi Prefecture Second People’s Hospital.}
\thanks{Jianqi Sun is also with the National Engineering Research Center of Advanced Magnetic Resonance Technologies for Diagnosis and Therapy (NERC-AMRT), and Med-X Research Institute, Shanghai Jiao Tong University, Shanghai, China.}
\thanks{Kun Wang is with the Renji Hospital, Shanghai, China}
\thanks{Jun Xu is with the Shanghai Sixth People’s Hospital, Shanghai, China}
\thanks{Corresponding authors: Jun Xu(e-mail: junxu19781214@163.com) and Jianqi Sun(e-mail: milesun@sjtu.edu.cn).}
}

\maketitle

\begin{abstract}
Osteoporotic vertebral compression fractures (OVCFs) are prevalent in the elderly population, typically assessed on computed tomography (CT) scans by evaluating vertebral height loss. This assessment helps determine the fracture's impact on spinal stability and the need for surgical intervention. However, the absence of pre-fracture CT scans and standardized vertebral references leads to measurement errors and inter-observer variability, while irregular compression patterns further challenge the precise grading of fracture severity.
While deep learning methods have shown promise in aiding OVCFs screening, they often lack interpretability and sufficient sensitivity, limiting their clinical applicability. 
To address these challenges, we introduce a novel vertebra synthesis-height loss quantification-OVCFs grading framework. Our proposed model, HealthiVert-GAN\footnote{Our code is available: \url{https://github.com/zhibaishouheilab/HealthiVert-GAN}.}, utilizes a coarse-to-fine synthesis network designed to generate pseudo-healthy vertebral images that simulate the pre-fracture state of fractured vertebrae. This model integrates three auxiliary modules that leverage the morphology and height information of adjacent healthy vertebrae to ensure anatomical consistency. Additionally, we introduce the Relative Height Loss of Vertebrae (RHLV) as a quantification metric, which divides each vertebra into three sections to measure height loss between pre-fracture and post-fracture states, followed by fracture severity classification using a Support Vector Machine (SVM). Our approach achieves state-of-the-art classification performance on both the Verse2019 dataset and in-house dataset, and it provides cross-sectional distribution maps of vertebral height loss. This practical tool enhances diagnostic accuracy in clinical settings and assisting in surgical decision-making.
\end{abstract}


\section{INTRODUCTION}
\label{sec:introduction}
\IEEEPARstart{O}{steoporotic} vertebral compression fractures (OVCFs) present significant clinical challenges, particularly affecting an aging population susceptible to osteoporosis and metastatic disease\cite{kanis2013european}. It is estimated that around 25\% women over 50 years old have at least one compression fracture. These fractures, resulting from the collapse of vertebral bodies under compressive stress, lead to chronic pain, impaired mobility, and diminished quality of life\cite{yu2018risk}. Furthermore, the occurrence of the OVCF may cause additional compression fractures within adjacent vertebrae\cite{klotzbuecher2000patients}. Early diagnosis and clear assessment of surgical need are therefore critical for OVCF patients.

Accurate grading and quantification of vertebral fractures play a crucial role in clinical decision-making, aiding surgeons in evaluating fracture severity and its impact on spinal stability to determine whether surgical intervention is necessary. The Genant semi-quantitative grading system, proposed by Genant et al.\cite{genant1993vertebral}, is a valuable diagnostic tool for OVCFs; however, it was initially designed for analysis based on standard lateral X-ray images under limited conditions at the time. Today, computed tomography (CT) has become the gold standard in OVCF evaluation as a three-dimensional imaging technique, widely adopted for assessing vertebral compression in multiple planes and offering improved accuracy. Clinical experience has shown that applying the Genant grading system to individual CT cross-sections can effectively assess compression severity\cite{leboff2022clinician}. However, current methods lack the capability to continuously analyze compression across multiple cross-sections, potentially overlooking severely compressed areas and misleading treatment.
Accurate OVCF grading is critical as it directly influences clinical intervention choices: normal and mild fractures typically require only conservative treatment, whereas moderate and severe fractures necessitate surgical intervention.
Therefore, it is essential to develop a new assessment method that analyzes vertebral fractures across multiple cross-sections or in a three-dimensional context.%, enabling a comprehensive evaluation of fracture stability and the necessity for surgical intervention.

Furthermore, current OVCFs grading processes still largely rely on visual assessment and manual measurement by surgeons. For junior surgeons, the lack of standardized pre-fracture vertebral references introduces measurement error, inter-observer variability, and potential misdiagnoses\cite{mitchell2017reporting}. For example, subtle, less visible fractures may go undetected. This highlights the need for automated, fully quantifiable methods, such as computer-aided diagnosis (CAD) systems\cite{kolanu2020clinical}, to support surgeons by improving the consistency, accuracy, and sensitivity in OVCF assessment.

\begin{figure}[!t]
\centerline{\includegraphics[width=\columnwidth]{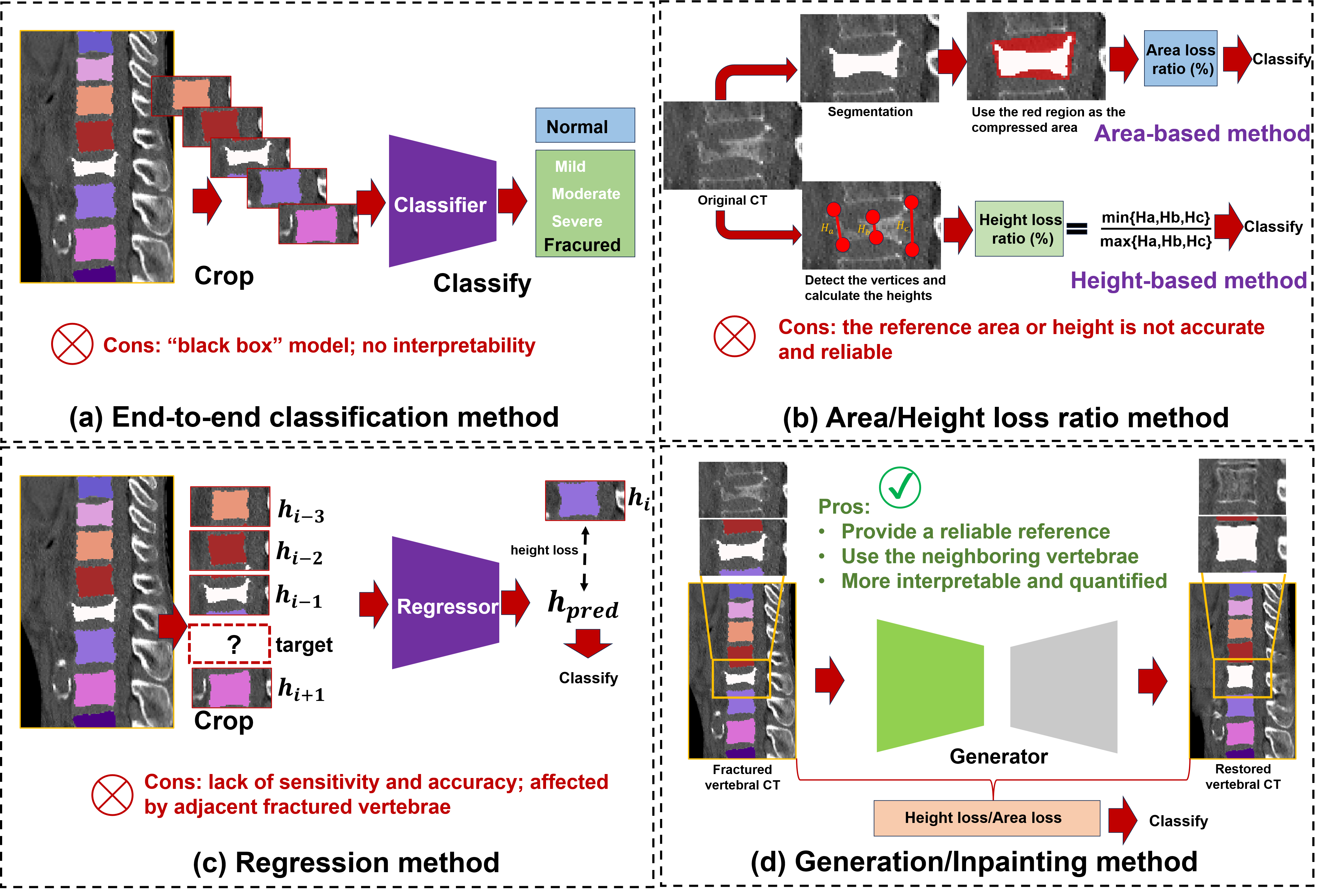}}
\caption{Illustrations of different OVCFs grading and quantification methods:(a) End-to-end classification methods, (b) Area/height loss ratio methods, (c) Regression methods, and (d) Generation/Inpainting methods.}
\label{comparison_with_others}
\end{figure}

Recently, deep learning (DL) models have been widely applied for the automatic screening and grading of OVCFs\cite{li2021differential,wang2024transformer}. These models are trained to recognize predictive features from large datasets of labeled radiographic images. Previous studies have explored three main types of models for the diagnosis and quantification of OVCFs (see \Cref{comparison_with_others}): (a) End-to-end classification methods; (b) Area/Height loss ratio methods and (c) Regression methods. Although these models have demonstrated their potential in vertebral fracture grading tasks, each presents certain limitations. Classification-based models typically adopt an end-to-end approach, classifying vertebrae images as either fractured or non-fractured or categorizing them according to the Genant grading system. However, classification-based models face notable challenges: (1) they often function as ``black-box'' models, which lack interpretability and fail to provide quantitative measures of compression, limiting clinical insight; (2) the long-tail effect in medical imaging datasets, where normal vertebrae far outnumber fractured ones, reduces classification accuracy. This imbalance makes it especially challenging to distinguish less common fracture cases.

To improve the interpretability of classification-based models, recent studies\cite{zakharov2023interpretable,shen2023using} have introduced area/height loss ratio metrics. However, these approaches face a fundamental limitation: the absence of true pre-fracture CT scans in clinical practice, as patients typically undergo imaging only after symptom onset.
Consequently, existing methods have to estimate the pre-fracture vertebral reference. These estimated vertebral heights or areas are often inferred from the post-fracture vertebral imaging, making them potentially inaccurate and unreliable. Other studies\cite{flanders2023height} have employed regression models, using the heights of adjacent vertebrae in CT images to predict the height of the target vertebra. However, if the adjacent vertebrae are fractured, the predicted height can be biased. More importantly, these methods often rely solely on two-dimensional X-rays or a single central CT slice. As a result, they cannot capture the height loss distribution across multiple sections, limiting the ability to analyze the fracture morphology. %Furthermore, single-slice quantification has lower sensitivity and accuracy compared to comprehensive three-dimensional analysis.

To address these issues, we propose the use of Generative Adversarial Networks (GANs)\cite{goodfellow2014generative} to restore and generate the pre-fracture state of vertebrae, providing three-dimensional pseudo-healthy vertebrae as references for quantifying height loss distribution, grading compression severity, and determining surgical indications for OVCFs. GANs have been extensively applied in medical image analysis, demonstrating remarkable success in generating realistic images\cite{yi2019generative,dayarathna2023deep}.  For an individual, the structural consistency of vertebrae and the predictable variation in vertebral heights \cite{frobin1997precision, verlaan2005three, frobin2002vertebral} enable the generation model to accurately predict healthy heights and shapes from adjacent vertebrae.
Several studies have utilized GANs for vertebrae generation task\cite{bukas2021patient,miao2019spinal} and validated their feasibility.

In this paper, we propose a novel three-step workflow for OVCF grading: (1) synthesis of pseudo-healthy vertebrae, (2) quantification of vertebral height loss, and (3) grading of compression severity. Firstly, we design a GAN-based synthesis network, named HealthiVert-GAN, for the synthesis of the pseudo-healthy vertebrae. HealthiVert-GAN utilizes a coarse-to-fine generation process, leveraging sagittal spine CT information, to generate volumetric vertebral images. To leverage contextual information from adjacent healthy vertebrae and reduce the impact of fractured vertebrae, we incorporate a HealthiVert-Guided Attention Module (HGAM) that focuses on non-fractured regions in the image. To better distinguish between the healthy and fractured vertebrae, we add an Edge-Enhancing Module (EEM) to capture the precise edge morphology. We also substitute the conventional mask-and-inpaint strategy with a Self-adaptive Height Restoration Module (SHRM), ensuring that the generated vertebrae meets individualized height standard of the healthy state. Together, these three modules enable the generated vertebrae to closely resemble a healthy state, providing a reliable reference for vertebral height loss quantification. Secondly, to assess the degree of compression in an interpretable and sensitive way, we introduce a new quantification metric, the Relative Height Loss of Vertebrae (RHLV), measured at the anterior, middle, and posterior segments of vertebrae. Lastly, these three RHLV values are sent into a Support Vector Machine (SVM) to classify the fracture level. %This methodology enhances diagnostic precision and offers actionable insights for clinical decision-making, significantly improving the utility of VCF assessments.

The main contributions of this paper are summarized as follows:

\begin{itemize}
    \item We introduce a novel framework for generating three-dimensional pseudo-healthy vertebrae, which provides an accurate and reliable reference for vertebral compression quantification.
\end{itemize}

\begin{itemize}
    \item We develop a HealthiVert-Guided Attention Module that enhances the morphological accuracy of generated vertebrae, ensuring they closely resemble healthy vertebrae within the image. 
\end{itemize}

\begin{itemize}
    \item We design a Self-adaptive Height Restoration Module that adaptively adjusts the height of the inpainted vertebrae, independent of the original vertebral shape and mask size, while aligning with the height variations of adjacent vertebrae.
\end{itemize}

\begin{itemize}
    \item We obtain distribution maps of vertebral height loss across multiple cross-sections and propose a new quantification metric. Based on the quantification results, our method achieves state-of-the-art performance in OVCF grading on two datasets and provides insights for assessing fracture stability and surgical indications.
\end{itemize}

\section{Related Work}

\subsection{Classification And Quantification Of Osteoporotic Vertebrae Compression}
OVCFs are commonly assessed using the semi-quantitative Genant grading system \cite{genant1993vertebral,burns2017vertebral,loffler2020vertebral,genant2000vertebral}, which classifies vertebrae based on relative height loss into four categories: normal (\textless20$\%$), mild (20$\%$-25$\%$) moderate (26$\%$-40$\%$), and severe (\textgreater40$\%$). While this system is widely used, it relies on manual measurements and lacks a standardized reference for healthy vertebral height, complicating accurate severity assessment. Attempts to use standard vertebral models for comparison have been limited by anatomical variations. Thomas et al. \cite{baum2014automatic} improved accuracy by segmenting vertebrae to calculate height ratios, but this method can still misclassify uniformly compressed vertebrae. Further research \cite{burns2017vertebral} suggested using height ratios relative to adjacent vertebrae, yet accuracy decreases in cases with multiple continuous fractures.

\subsection{Deep Learning In Vertebrae Fracture Grading and Quantification}
Deep learning (DL) models have been applied to automate OVCF screening and grading. Dong et al. \cite{dong2022deep} used GoogleNet to classify fracture levels, while Nicolaes et al. \cite{nicolaes2020detection} employed 3D CNNs for fracture detection in CT scans. Xin et al. \cite{wei2022faint} enhanced accuracy with self-supervised contrastive learning. However, these end-to-end models often lack interpretability, a critical issue for clinical decision-making. Recent efforts have integrated quantifiable metrics with DL. Alexey et al. \cite{zakharov2023interpretable} and Li et al. \cite{shen2023using} proposed height and area loss ratios, but these methods rely on single-slice analysis and lack a reliable pre-fracture reference, limiting their accuracy and consistency.

\subsection{Generative Adversarial Networks In Medical Image Synthesis}
GANs have shown significant promise in medical imaging\cite{yi2019generative,dayarathna2023deep}, particularly in image inpainting \cite{wei2020slir,yu2020deep}. Recent studies have applied GANs to vertebral inpainting tasks. Christina et al. \cite{bukas2021patient} used GANs to replace fractured vertebrae for volume estimation in vertebroplasty planning, while Miao et al. \cite{miao2019spinal} generated normal vertebral images from tumor-affected CT scans. However, these approaches face challenges in cases with multiple continuous fractures and are limited by conventional mask strategies that restrict generation space. Despite these limitations, GANs offer a promising avenue for providing reliable references for healthy vertebral morphology, crucial for accurate fracture assessment.

\begin{figure*}[!t]
\centerline{\includegraphics[width=1.6\columnwidth]{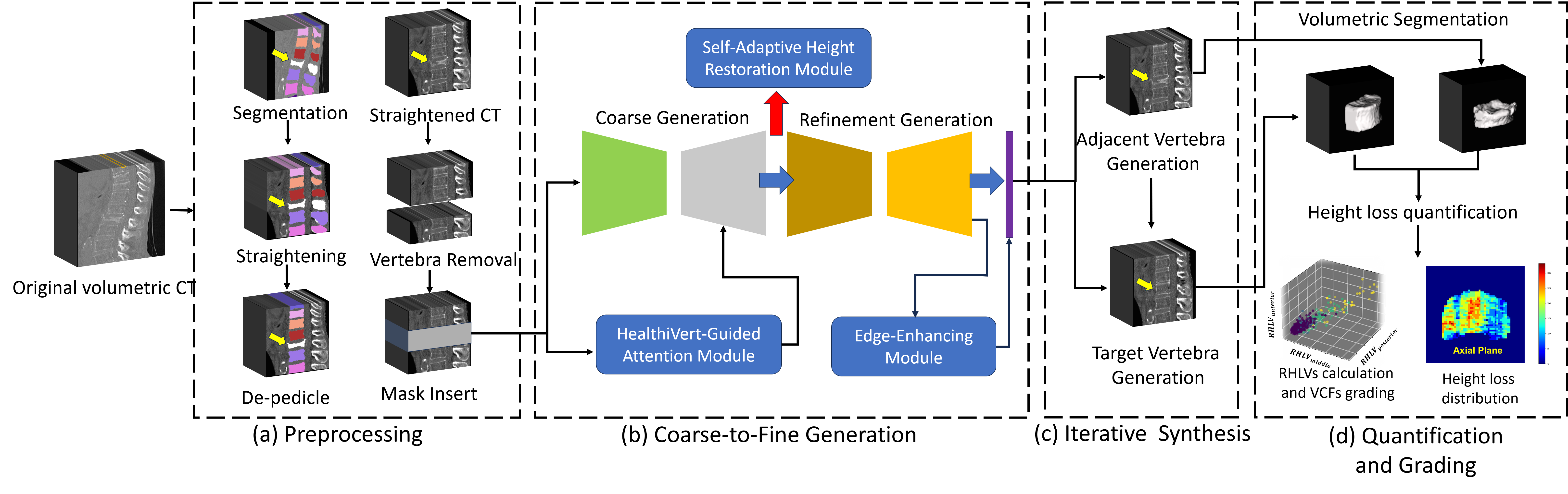}}
\caption{Workflow of our proposed method, including four main phases: (a) preprocssing, (b) synthesis, (c) quantification, and (d) classification.}
\label{workflow}
\end{figure*}

\section{PROPOSED METHOD}
As shown in \Cref{workflow}, our approach to OVCF grading is divided into four main phases: A. vertebral preprocessing; B. two-stage generation framework; C. iterative synthesis; and D. quantification and fracture grading.

\subsection{Vertebral Preprocessing}
\subsubsection{Segment the vertebrae and straighten the spine}
Segmentation of vertebrae is used for further analysis and quantification. In this work, we adopt the state-of-the-art vertebrae segmentation model, SCNet\cite{payer2020coarse}, which is the champion method in the MICCAI 2019 Large Scale Vertebrae Segmentation Challenge. After segmentation, we conduct spine straightening for both CT and segmentation results. Proper alignment of vertebrae in a vertical configuration is essential for accurately comparing vertebral height. Utilizing the methodology developed by Alexey et al.\cite{zakharov2023interpretable}, we straighten the spine in three-dimensional space.

\subsubsection{De-pedicle}
De-pedicle is then applied to the segmentation results. Specifically, we remove the vertebral arch in the segmentation, as vertebral height measurements are based solely on the vertebral body. By eliminating the pedicles, generative models can focus on reconstructing and analyzing the vertebral body separately. Operationally, the process begins with ensuring the spine is correctly straightened to properly align the vertebrae. Next, the layer in the coronal plane where the vertebra visibly splits, indicating the start of the vertebral arch, is identified. From the identified starting layer, all posterior layers that constitute the pedicles are removed in the segmentation.

\subsubsection{Remove and mask the vertebra}
To eliminate the influence of the original vertebrae on the synthesis process, we employ a novel mask strategy for CT inpainting. This involves designing a mask that completely covers the target vertebrae and removes all height cues. Initially, the upper and lower coordinates ($x_{upper}$ and $x_{lower}$) of the target vertebra are determined based on its segmentation. We define the real vertebral height as: $h_{real} = x_{lower}-x_{upper}$. Subsequently, the CT image is segmented into three parts based on $x_{upper}$ and $x_{lower}$: $CT_{upper}$, $CT_{vert}$, and $CT_{lower}$. The $CT_{vert}$ part, which contains the target vertebra, is removed to ensure that no residual characteristics influences the new vertebral image synthesis.
A zero-value mask, set to a fixed height of $h_{max}$, is then inserted between the $CT_{upper}$ and $CT_{lower}$ parts. $h_{max}$, determined as 40mm based on clinical insights that normal vertebral heights rarely exceed this value, establishes a boundary for the maximum height of the generated vertebra. This process is depicted in \Cref{SHRM_and_HGAM} (b), Step 1. This setup allows for the height of the generated vertebra to be unrestricted by the original compressed state, while remaining adaptable within a reasonable range. The precise height adjustments are managed by our proposed Self-adaptive Height Restoration (SHRM) Module (described in the next section).

\begin{figure*}[ht]
\begin{center}
\includegraphics[width=1.6\columnwidth]{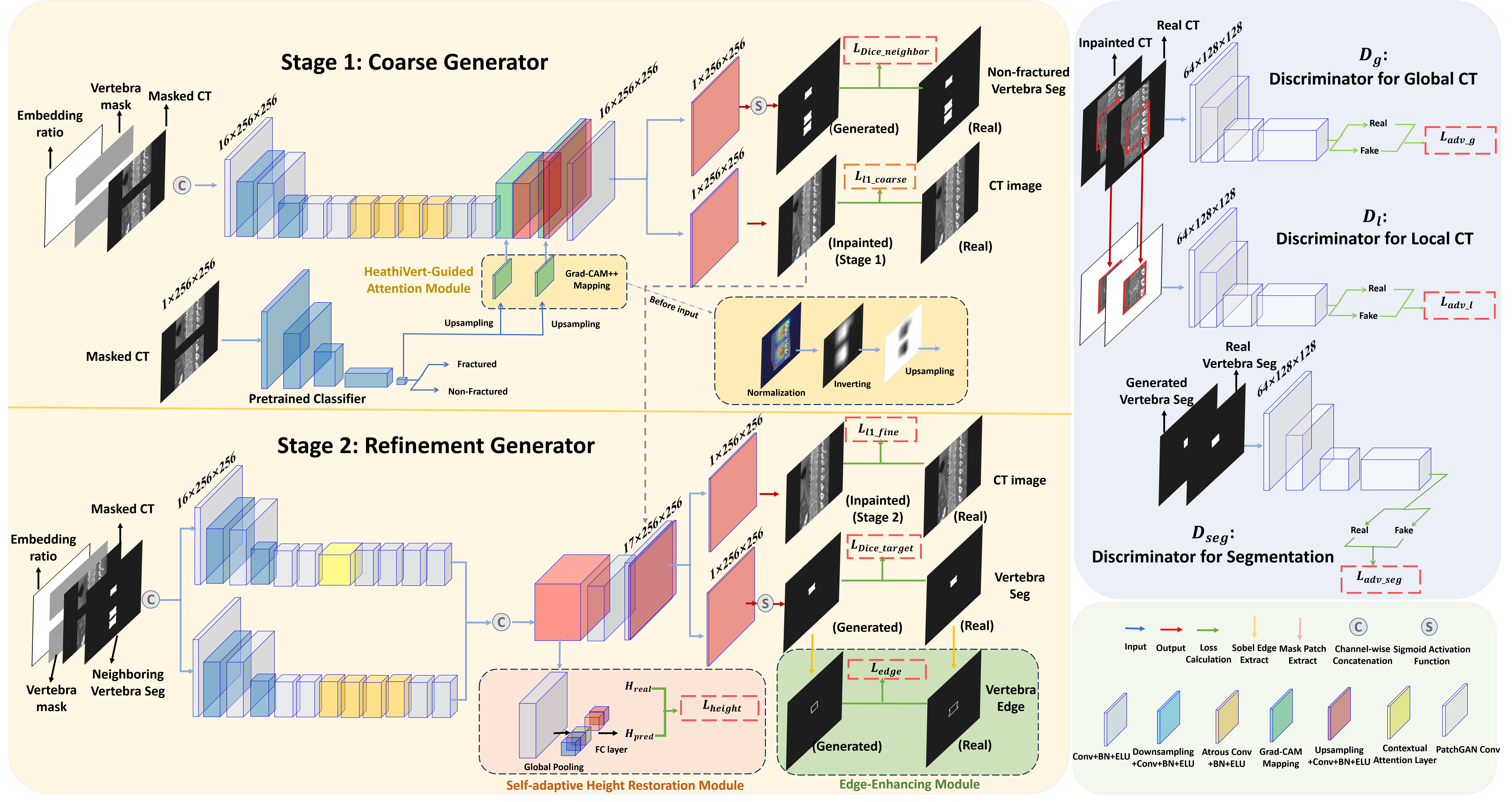}
\end{center}
\caption{Detailed architecture of HealthiVert-GAN, including two-stage generation and three proposed modules named HGAM, EEM, and SHRM.}
\label{network}
\end{figure*}

\subsection{HealthiVert-GAN: a Two-Stage Vertebral CT Generation GAN}
As shown in \Cref{network}, the HealthiVert-GAN model leverages a two-stage inpainting approach to generate these pseudo-healthy vertebrae, in a coarse-to-fine way.
\subsubsection{Coarse-to-Fine Generation}
This architecture adopts a sequential two-stage generation processes: a coarse stage followed by a refinement stage. To achieve our goals of generating pseudo-healthy vertebrae and learning from the adjacent vertebrae, we have incorporated three innovative modules into the architecture.
\begin{itemize}
    \item \emph{Coarse generation:} The synthesis begins with a coarse generator featuring an U-like architecture. The input to the coarse generator consists of three components: masked CT image, corresponding mask image that delineates the generation boundary, and normalized position ratio, which indicates the position of the slice in the volumetric CT. These three inputs are concatenated along the channel dimension, forming a $3\times256\times256$ feature map. The coarse generator performs two tasks: the generation task of generating a coarse pseudo-healthy vertebral CT image, referred to as the coarse CT, and the segmentation task of segmenting neighboring non-fractured vertebrae in the masked CT. The coarse CT provides auxiliary information to the subsequent refinement generation, while the segmentation of neighboring non-fractured vertebrae helps generator implicitly capture the essential features of adjacent vertebrae in the image, especially the non-fractured characteristics. The task of Adjacent Healthy Vertebrae Segmentation is referred as AHVS in our paper. %By incorporating the segmentation branch, the generator not only integrates contextual anatomical features but also captures long-range dependencies across the vertebrae, enhancing the generation quality.
\end{itemize}

\begin{itemize}
    \item \emph{Refinement generation:} The four-channel input feeds into the fine generator, comprising the masked CT, corresponding mask, normalized position ratio, and the non-fractured vertebrae segmentation. The architecture features two encoder branches: one mirroring the coarse generator to maintain continuity, and another enhanced with a contextual attention mechanism\cite{yu2018generative} to improve details. The outputs of two encoders are concatenated and then upsampled. The coarse CT is incorporated into the final upsampling layers to refine the synthesized image details. Finally, this refined generator produces both the refined CT image and the binary segmentation of the targeted vertebra. The segmentation output is essential, aiding the focus on precise reconstruction of the target area and directly facilitating subsequent measurements of vertebral height compression.
\end{itemize}

\subsubsection{HealthiVert-Guided Attention Module}
During the coarse generation stage, we incorporate a segmentation task of non-compressed vertebrae in the image (AHVS). However, this segmentation task alone may inadvertently include fractured vertebrae, as the segmentation branch lacks the capability to distinguish between fractured and healthy vertebrae, potentially incorporating characteristics from adjacent fractured vertebrae.

\begin{figure}[ht]
\begin{center}
\includegraphics[width=\columnwidth]{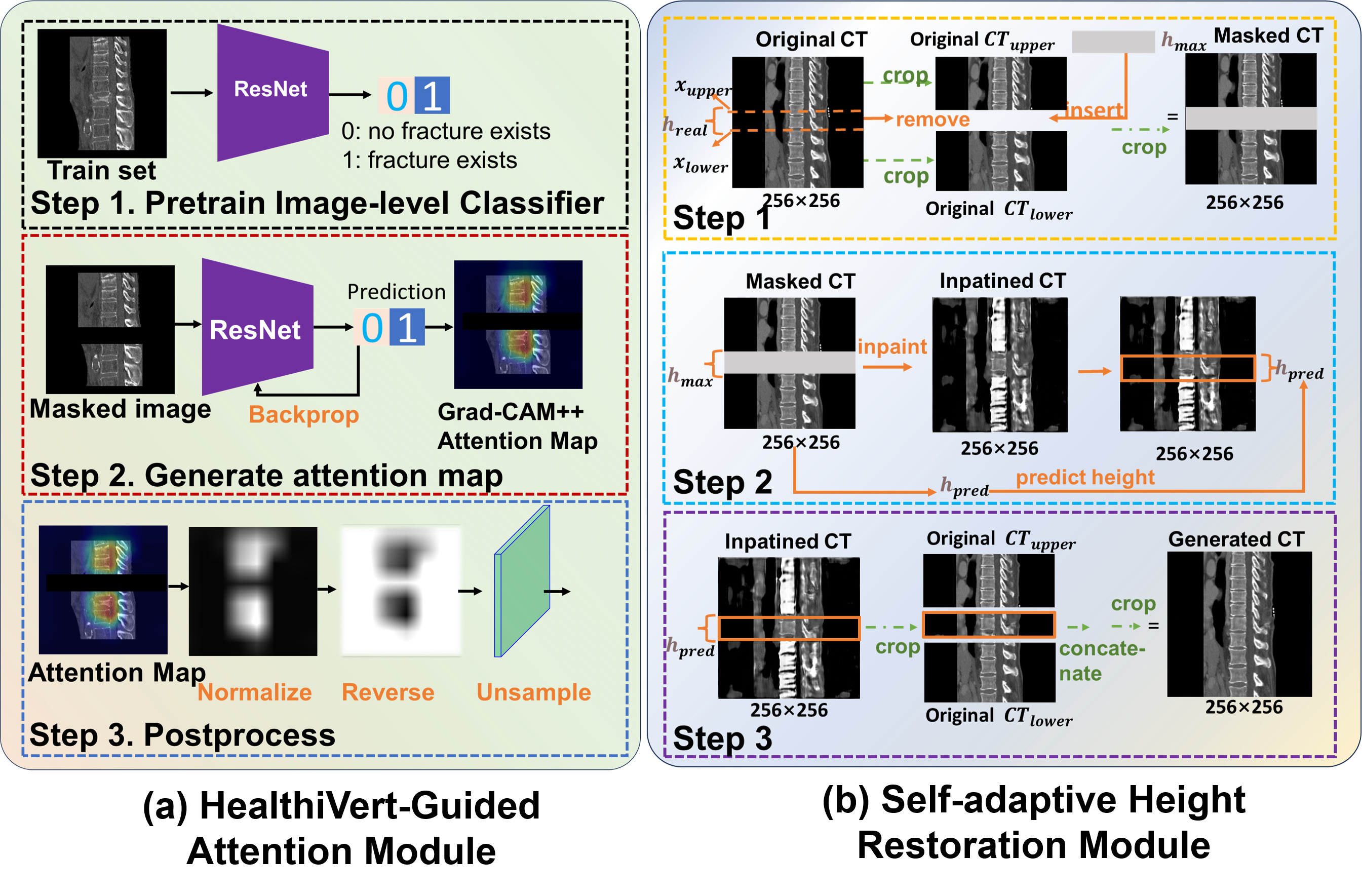}
\end{center}
\caption{(a) HealthiVert-Guided Attention Module helps model to distinguish the fractured and non-fractured vertebrae in the image, minimizing the influence of adjacent fractures. Potential fractured areas are highlighted in the attention map. (b) Self-adaptive Height Restoration Module helps the generated vertebra restore to its predicted healthy height in a self-adaptive way. Step 1 involves removing the target vertebra and inserting a fixed-height mask. In Step 2, the masked area is inpainted, and the pre-fracture vertebral height $h_{pred}$; Step 3 crops the inpainted CT based on $h_{pred}$, obtaining the final generated CT. }
\label{SHRM_and_HGAM}
\end{figure}

To address this issue, we introduce the HealthiVert-Guided Attention Module (HGAM, as shown in \Cref{SHRM_and_HGAM} (a)). Initially, we employ a pretrained image-level classification model\cite{wei2022faint} that categorizes CT images of vertebrae into two classes: positive, indicating compression fractures are present, and negative otherwise. After pre-training, we input the masked CT images into this pretrained network, utilizing the Grad-CAM++\cite{chattopadhay2018grad} technique to generate attention maps at the final output layer of the classification model.
These Grad-CAM++ maps highlight the network’s focal areas during classification, identifying potential fracture regions with elevated values. By converting these values to a range between 0 and 1 and then inverting them, areas without fractures are emphasized, and fracture regions are diminished. We resize these maps to match specific dimensions ($128 \times 128$ and $256 \times 256$) and concatenate them to the corresponding layers of the coarse generator. This integration helps the model focus on healthy vertebrae, ensuring precise segmentation of non-compressed vertebrae and minimizing the influence of adjacent fractures on the synthesized outcomes.

\subsubsection{Edge-Enhancing Module}
In distinguishing between non-fractured and fractured vertebrae, precise edge morphology is essential. Traditional segmentation tasks, while effective for general structure delineation, often fail to capture the finer edge details that are critical for identifying subtle morphological changes associated with fractures. These changes, such as slight indentations and localized concavities, are essential indicators of minor fractures. These morphological features must be accurately handled in synthetic images to ensure they represent healthy vertebrae for effective diagnostic use.
To address this, we incorporate an Edge-Enhancing Module into our generative model, following the fine generation stage. We utilize a Sobel edge extractor to produce an edge map from the target vertebra segmentation. An edge loss function is introduced that computes the Mean Square Error (MSE) between the edge maps of the generated and actual images. By penalizing discrepancies in these edge maps, the model is compelled to improve its accuracy in replicating detailed edge features, essential for effective diagnosis. %The effectiveness of precise edge representation in image synthesis has been validated by Li et al.\cite{li2023multi}, who employed a Sobel edge extractor and pretrained their encoder to extract high-level features for improved performance in tasks like morphology classification and lesion segmentation.

\subsubsection{Self-Adaptive Height Restoration Module}
Conventional image inpainting typically uses a mask stragety that roughly covers the target object in CT images, limiting image generation to the masked region and preserving the rest of the image unchanged. If the mask is designed based on the size and shape of the target compressed vertebra, the generated vertebral height cannot fully restore to its pre-fractured state, especially in the condition that the target vertebra is severely compressed.
In our approach, we introduce a Self-Adaptive Height Restoration Module (SHRM, as shown in \Cref{SHRM_and_HGAM} (b)) that adaptively adjusts to the predicted pre-fractured vertebral height $h_{pred}$. This module integrates a branch before the generator decoder, which outputs the value of the predicted height. The generated vertebral image within the fixed-height $h_{max}$ mask is then cropped to this predicted height $h_{pred}$ and seamlessly inserted between original $CT_{upper}$ and $CT_{lower}$. This method, while straightforward, is crucial for accurately restoring the vertebral height.

\subsubsection{Loss functions}
The losses of our proposed generator consist of five parts: (1) $\ell_1$ loss, (2) adversarial loss, (3) Dice loss, (4) edge loss, and (5) height loss. $\ell_1$ loss is defined as follows:
\begin{equation}\label{eq1}
	\begin{split}
	\mathcal{L}_{\ell_1} =&\lambda_1\cdot||CT_{real}-G_{coarse}(CT_{masked})||_1\\
    &+\lambda_2\cdot||CT_{real}-G_{fine}(CT_{masked})||_1 
	\end{split}
\end{equation}

Here, $G_{coarse}(CT_{masked})$ is the generated CT from the coarse generator and $G_{fine}$ is from the refinement generator. $\lambda_1$ and $\lambda_2$ are both set to 0.5. It should be noted that $\mathcal{L}_{\ell_1}$ loss is normalized by the mask size to ensure proper scaling. 

We design three discriminators, each being a convolutional PatchGAN\cite{isola2018imagetoimage} classifier offering an adversarial loss. $D_g$ is global for the global CT image and $D_l$ is local for masked region of the CT image, and $D_{seg}$ is for the segmentation. The corresponding adversarial losses are defined as follows:
\begin{equation}\label{eq2}
\begin{split}
\mathcal{L}_{adv_g}& = \mathbb{E}_{(CT_{real})}[\log 
D_g(CT_{real})] \\
&+ \mathbb{E}_{(CT_{masked})}[\log (1 - D_g(G_{fine}(CT_{masked}))]
\end{split}
\end{equation}
\begin{equation}\label{eq3}
	\begin{split}
    \mathcal{L}_{adv_l}& = \mathbb{E}_{(CT_{real})}[\log D_l(CT_{real}^{local})] \\
    &+ \mathbb{E}_{(CT_{masked})}[\log (1 - D_l(G_{fine}^{local}(CT_{masked}))]
	\end{split}
\end{equation}
\begin{equation}\label{eq4}
	\begin{split}
    \mathcal{L}_{adv_seg}& = \mathbb{E}_{(Seg_{real})}[\log D_{seg}(Seg_{real})] \\
    &+ \mathbb{E}_{(CT_{masked})}[\log (1 - D_{seg}(G_{seg}(CT_{masked}))]
	\end{split}
\end{equation}

Here, $CT_{real}^{local}$ and $G_{fine}^{local}$ indicate the masked region of the real CT and the generated images. $G_{seg}$ indicates the segmentation output from the refinement generation.

The overall adversarial loss is defined as follows:
\begin{equation}\label{eq5}
\mathcal{L}_{adv}=\lambda_{adv1}\cdot\mathcal{L}_{adv_g}+\lambda_{adv2}\cdot\mathcal{L}_{adv_l}+\lambda_{adv3}\cdot\mathcal{L}_{adv_seg}
\end{equation}

Here, $\lambda_{adv1}$, $\lambda_{adv2}$, and $\lambda_{adv3}$ are set to $\frac{1}{3}$ in our experiments.

Dice loss consists of two parts: the Dice loss for non-fractured vertebrae segmentation and for the targeted vertebra segmentation. We average these two to get the overall Dice loss $\mathcal{L}_{Dice}$.

Edge loss is formulated as follows:
\begin{equation}\label{eq6}
\mathcal{L}_{edge} = ||\mathcal{S}(Seg_{real})-\mathcal{S}(G_{seg}(CT_{masked}))||_1
\end{equation}

Here, $\mathcal{S}$ indicates the Sobel edge extractor to obtain the edge mapping of the segmented vertebrae.

Height loss is calculated as follows:
\begin{equation}\label{eq7}
    \mathcal{L}_{height} = \lambda_{h_1}\cdot\frac{|h_{real}-h_{pred}^{coarse}|}{h_{real}}
    + \lambda_{h_2}\cdot\frac{|h_{real}-h_{pred}^{fine}|}{h_{real}}
\end{equation}

Here, $h_{pred}^{coarse}$ and $h_{pred}^{fine}$ indicate the predicted heights in the coarse and refinement generation, respectively, using the proposed Self-adaptive Height Restoration Module. $h_{real}$ is the actual height of the target masked vertebra. We set $\lambda_{h_1}$ and $\lambda_{h_2}$ to 0.5 in our experiments.

Our overall loss function is defined as follows:
\begin{equation}\label{eq8}
    \begin{split}
    \mathcal{L}_{overall} = &\lambda_{\ell_1}\cdot\mathcal{L}_{\ell_1}+\lambda_{adv}\cdot\mathcal{L}_{adv}+\lambda_{Dice}\cdot\mathcal{L}_{Dice}\\
    &+\lambda_{edge}\cdot\mathcal{L}_{edge}+\lambda_{height}\cdot\mathcal{L}_{height}
    \end{split}
\end{equation}

For our experiments, we empirically choose $\lambda_{adv}=1$, $\lambda_{Dice}=20$, $\lambda_{\ell_1}=40$, $\lambda_{height}=80$, $\lambda_{edge}=80$ for training.

\subsection{Two-step Iterative Synthesis}
The two-step iterative synthesis begins by generating adjacent vertebrae, followed by the iterative generation of the target vertebra. Subsequently, images from sagittal planes are stacked to create a comprehensive 3D depiction of the vertebrae.

\subsubsection{Iterative synthesis}
This step is used to further optimized the generation. In cases where compression fractures are prevalent across multiple vertebrae, a single-step generation process may fail to accurately simulate non-fractured states due to the influence of the adjacent fractures. This is common in OVCF patients because one vertebra fracture often leads to continuous adjacent fractures. To enhance the reliability of generated vertebral images, we implement a two-step iterative generation process. In the first step, the vertebrae adjacent to the target vertebra are restored using the original $CT_{image}$. This initial synthesized CT is labeled as $CT_{step1}$. In the second step, we mask the target vertebra and generate its non-fractured state based on $CT_{step1}$. This iterative strategy ensures that the adjacent vertebrae of the target one are not severely compressed. By doing so, it decreases the probability that the target vertebra will exhibit characteristics associated with fracture states, which is particularly useful in cases involving heavily compressed spinal segments. 

\subsubsection{3D construction from sagittal plane}
We employ a 2D generative model based on HealthiVert-GAN and process all layers of the target vertebra from the sagittal view, as surgeons typically use the sagittal plane to diagnose vertebral fractures. We input the normalized position ratio, which indicates the relative position of each 2D slice in the volume, to the generator. The \textit{i-th} slice ratio value $ratio_i$ is calculated as:
\begin{equation}
ratio_i = \frac{|2i-N|}{N}
\end{equation}
where \textit{N} is the number of sagittal slices. The generation outputs of all slices are then stacked to create volumetric data. To investigate the generation performance of the sagittal and coronal planes of the images, we run two separate 2D models and compare the results in the experiments section.

\subsection{RHLV Calculation and Fracture Grading}
Based on the following synthesis results, we calculate the height difference between real vertebrae and the generated ones, introducing a new quantification metric: Relative Height Loss of Vertebrae (RHLV). For the vertebrae grading, we construct a classier using SVM and input the RHLVs to grade the vertebral compression severity.

The RHLV is defined as shown in \Cref{eq9}, which quantifies the compression level by comparing the average heights of the pseudo-healthy generated vertebra ($H_{gen}$) to the original vertebra ($H_{ori}$). We calculate the average heights using the 3D volumetric vertebral body rather than just the center slices. %This method ensures a more sensitive and reliable assessment than 2D methods, particularly in cases where the compression affects areas beyond the center of the vertebra.
\begin{equation}\label{eq9}
    RHLV = \frac{H_{gen}-H_{ori}}{H_{gen}}
\end{equation}

There are different types of compression fractures, varying in how they impact different parts of the vertebra: biconcave fractures typically affect the center, wedge fractures impact the anterior, and crush fractures affect the posterior. To address this variability, we uniformly segment each vertebra into three volumetric regions along the sagittal axis, as anterior, middle, and posterior. We calculate RHLV for each region, including $ RHLV_{anterior}$, $ RHLV_{middle}$, and $ RHLV_{posterior}$, to ensure that our assessment accurately captures the detailed structural changes in the vertebra, enhancing the sensitivity and accuracy of our fracture grading system. This segmented approach allows us to detect subtle variations in height loss across different parts of the vertebra.

To grade fractures based on RHLV, we have constructed a SVM classifier. $ RHLV_{anterior}$, $ RHLV_{middle}$, and $ RHLV_{posterior}$, along with the corresponding vertebral compression fracture labels, are input into the SVM for training. %Also, the quantification results and the height loss distribution mapping can be used in the clinical guidance.

\begin{table}[ht]
\centering
\caption{Characteristics of CT Scans and Patients in the Verse 2019 and In-house Dataset}
\label{tab:datasets}
\begin{tabular}{@{}lcc@{}}
\toprule
\textbf{Category} & \textbf{Verse 2019\cite{loffler2020vertebral}} & \textbf{In-house} \\ \midrule
\multicolumn{3}{l}{\textbf{Patients}} \\ \midrule
No. of patients & 141 & 429 \\
No. of women & 92 & 311 \\
Age (yr*) & 66.1±15 & 70.0±10 \\ \midrule
\multicolumn{3}{l}{\textbf{Imaging}} \\ \midrule
No. of scans & 160 & 497 \\
No. of vertebrae & 1505 & 4325 \\
\quad Thoracic & 884 & 2308 \\
\quad Lumbar & 621 & 2017 \\ \midrule
\multicolumn{3}{l}{\textbf{Fractures}} \\ \midrule
No. of fractures (\% of total) & 288 (19.1\%) & 938 (21.7\%) \\
Grade 1 (mild, \% of total) & 141 (9.4\%) & 508 (11.7\%) \\
Grade 2 (moderate, \% of total) & 96 (6.3\%) & 290 (6.7\%) \\
Grade 3 (severe, \% of total) & 51 (3.4\%) & 140 (3.2\%) \\ \bottomrule
\multicolumn{3}{l}{* Data are means ± standard deviations} \\
\end{tabular}
\end{table}

\section{EXPERIMENTS AND RESULTS}
\subsection{Datasets}
We use two datasets in our experiment: Verse2019, a public dataset for fair comparison with other methods, and an in-house dataset we have collected for external validation to prove the generalization of our method. The details are described in \Cref{tab:datasets}.
\subsubsection{Verse2019}
The Verse2019 dataset \cite{loffler2020vertebral} comprises 160 scans from 141 patients, including detailed vertebral compression fracture labels annotated by radiologists using the Genant grading system, which categorizes vertebrae based on the degree of compression into four classes: non-fractured (1217 cases), mildly fractured (142 cases), moderately fractured (96 cases), and severely fractured (51 cases). Note that we only use the thoracic and lumbar vertebrae in the dataset.
\subsubsection{In-house dataset}
This retrospective study was approved by the Ethics Committee of Shanghai Sixth People’s Hospital (Approval No:2022-KY-101(K)). We collected patients from the Shanghai Sixth People’s Hospital between January 2019-January 2021. The inclusion criteria comprised: (1) Postmenopausal/perimenopausal women or males aged $>60$ years; (2) Documented diagnosis of osteoporosis and spinal fracture and were determined to be nonviolent fractures; (3) Availability of complete imaging series. Exclusion criteria were: (1) Imaging artifacts affecting diagnostic interpretation; (2) Secondary fractures from metastatic disease, spinal tuberculosis, high-impact trauma, osteoarthritic vertebral wedging, or endplatitis short vertebrae; (3) Prior vertebral augmentation therapy (e.g., percutaneous kyphoplasty) within 6 months. The final cohort comprises 497 CT scans from 429 patients. It comprises 508 mild fractures, 290 moderate fractures, 140 severe fractures, and 3387 normal vertebrae. We have divided the dataset into training, validation, and test sets in a 6:2:2 ratio, maintaining a consistent ratio of fracture categories within three subsets. 

The annotations were independently performed by a senior surgeon (J.X.) with over ten years of experience and a junior surgeon (C.C.) with approximately three years of experience. Both surgeons were trained in OVCF grading according to the Genant grading system and used the open-source software ITK-SNAP \cite{py06nimg} during annotation. Following quality assurance checks by another senior radiologist (K.W., with over ten years of experience), the senior surgeon's annotations were established as the reference standard.

\subsection{Implementation Details}
Our framework, HealthiVert-GAN, is implemented in the Pytorch framework, and all experiments are performed on a workstation equipped with an NVIDIA 3090 GPU. For optimization, we employ a batch size of 16 and adopt the Adam optimizer, in which $\beta_1$ is set to 0.5. All experiments are trained for 1000 epochs to ensure model convergence. The learning rate starts at 0.0002, and linearly decays after 100 epochs. The input image size for the generative model is set to $256 \times 256$. For CT images, we add an intensity window of [-300, 800]HU to highlight the vertebrae body. For each vertebra, we crop the image centered on the vertebral body, and then zero-pad it to $256 \times 256$. For the pre-training of the HGAM, we employ a 2D classifier and refer to the Wei et al.\cite{wei2022faint} for the detailed experiment settings. For the generation task, we only use the non-fractured cases in the dataset for training and validation. %Based on the trained generative model, we generate the pseudo-healthy images for all vertebrae and quantify the height loss.
For the classification task, we use five-fold cross-validation to evaluate the method, ensuring robustness to data splits. 
\begin{table*}[ht]
\centering
\caption{Ablation studies for preprocessing, modules, two-stage synthesis, and discriminators.}
\label{tab:ablation}
\begin{tabular}{|l|c|c|c|c|c|c|c|c|c|c|}
\hline
\multicolumn{11}{|c|}{\textbf{Ablation for preprocessing, AHVS, HGAM, and EEM}}\\ \hline
\multirow{2}{*}{\textbf{Method}} & \multicolumn{3}{c|}{\textbf{Modules}} & \multicolumn{4}{c|}{\textbf{Generation metrics}} & \multicolumn{3}{c|}{\textbf{Classification metrics}}\\
\cline{2-11}
& \textbf{AHVS} & \textbf{HGAM} & \textbf{EEM} & \textbf{PSNR$\uparrow$} & \textbf{SSIM$\uparrow$} & \textbf{Dice$\uparrow$} & \textbf{RHDR$\downarrow$} & \textbf{macro-P$\uparrow$} & \textbf{macro-R$\uparrow$} & \textbf{macro-F1$\uparrow$} \\
\hline
\multirow{5}{*}{w preprocessing} & $\times$ & $\times$ & $\times$ & 27.38& 0.892 & 0.875 & 10.58\% & 0.656 & 0.650 & 0.637  \\
& \checkmark & $\times$ & $\times$ & 27.41 & 0.896 & 0.883 & 7.89\% & 0.690 & 0.694 & 0.675  \\
& $\times$ & \checkmark & $\times$ & 27.70 & 0.907 &0.884 & 7.69\% & 0.675 & 0.673 & 0.658  \\
& \checkmark & \checkmark & $\times$ & 27.43 & 0.912 & 0.891 & 7.12\% & 0.701 & 0.723 & 0.688   \\
& \checkmark & \checkmark & \checkmark & \textbf{27.92} & \textbf{0.921} & \textbf{0.894} & \textbf{6.58\%} & \textbf{0.727} & \textbf{0.753} & \textbf{0.723}   \\
\hline
w/o straightening & \checkmark & \checkmark & \checkmark & 23.11 & 0.797 & 0.829 & 10.83\% & 0.518 & 0.586 & 0.524   \\
w/o de-pedicle & \checkmark & \checkmark & \checkmark & 25.31 & 0.863 & 0.823 & 7.20\% & 0.582 & 0.599 & 0.575   \\
\hline
\multicolumn{11}{|c|}{\textbf{Ablation for two-stage synthesis}}\\ \hline
\multirow{2}{*}{\textbf{Method}} & \multicolumn{3}{c|}{\textbf{Plane}} & \multicolumn{4}{c|}{\textbf{Generation metrics}} & \multicolumn{3}{c|}{\textbf{Classification metrics}}\\
\cline{2-11}
& \textbf{Sagittal} & \textbf{Coronal} & \textbf{Fusion*} & \textbf{PSNR$\uparrow$} & \textbf{SSIM$\uparrow$} & \textbf{Dice$\uparrow$} & \textbf{RHDR$\downarrow$} & \textbf{macro-P$\uparrow$} & \textbf{macro-R$\uparrow$} & \textbf{macro-F1$\uparrow$} \\
\hline
\multirow{3}{*}{one-stage} & \checkmark &  &  & 23.21 & 0.811 & 0.843 & 10.64\% & 0.591 & 0.585 & 0.574 \\
&  & \checkmark &  & 20.71 & 0.612 & 0.831 & 13.25\% & 0.578 & 0.613 & 0.589  \\
&  &  & \checkmark & - & - & - & - & 0.615 & 0.637 & 0.608   \\
\hline
\multirow{3}{*}{two-stage} & \checkmark &  & & \textbf{27.92} & \textbf{0.921} & \textbf{0.894} & \textbf{6.58\%} & \textbf{0.727} & \textbf{0.753} & \textbf{0.723}  \\
&  & \checkmark &  & 25.27 & 0.812 & 0.854 & 7.49\% & 0.666 & 0.730 & 0.669  \\
&  &  & \checkmark & - & - & - & - & 0.705 & 0.729 & 0.688  \\
\hline
\multicolumn{11}{|c|}{\textbf{Ablation for discriminators}}\\ \hline
\multirow{2}{*}{\textbf{Method}} & \multicolumn{3}{c|}{\textbf{Discriminator}} & \multicolumn{4}{c|}{\textbf{Generation metrics}} & \multicolumn{3}{c|}{\textbf{Classification metrics}}\\
\cline{2-11}
& \textbf{$D_g$} & \textbf{$D_l$} & \textbf{$D_{seg}$} & \textbf{PSNR$\uparrow$} & \textbf{SSIM$\uparrow$} & \textbf{Dice$\uparrow$} & \textbf{RHDR$\downarrow$} & \textbf{macro-P$\uparrow$} & \textbf{macro-R$\uparrow$} & \textbf{macro-F1$\uparrow$} \\
\hline
\multirow{4}{*}{w discriminator(s)} & \checkmark & $\times$ & $\times$ & 27.36 & 0.913 & 0.878 & 7.89\% & 0.679 & 0.703 & 0.677   \\
& $\times$& \checkmark & $\times$ & 27.67 & 0.911 & 0.883 & 7.64\% & 0.696 & 0.716 & 0.688 \\
& \checkmark & \checkmark & $\times$ & 27.71 & 0.915 & 0.891 & 6.58\% & 0.705 & 0.725 & 0.695  \\
& \checkmark & \checkmark & \checkmark & \textbf{27.92} & \textbf{0.921} & \textbf{0.894} & \textbf{6.44\%} & \textbf{0.727} & \textbf{0.753} & \textbf{0.723}  \\
\hline
\multicolumn{8}{l}{\small $\bullet$ Variances of classification for all experiments are less than 0.0001, so not displayed.} \\
\multicolumn{10}{l}{AHVS: Adjacent Healthy Vertebrae Segmentation; HGAM: HealthiVert-Guided Attention Mudule; EEM: Edge Enhancing Module} \\
\multicolumn{10}{l}{* Fusion refers to combining the quantification results from both views for classification.}
\end{tabular}
\end{table*}

\begin{table}[ht]
\centering
\caption{Evaluation of mask strategies, iterative synthesis, classifiers, and RHLVs}
\label{tab:evaluation}
\scalebox{0.80}{
\begin{tabular}{|c|c|c|c|c|c|}
\hline
\multicolumn{6}{|c|}{\textbf{Evaluation of mask strategies and iterative synthesis}}\\ \hline
\multirow{2}{*}{\textbf{Method}} & \multicolumn{2}{|c|}{\multirow{2}{*}{\textbf{Mask Strategy*}}} & \multicolumn{3}{c|}{\textbf{Classification metrics}}\\
\cline{4-6}
& \multicolumn{2}{|c|}{} & \textbf{macro-P$\uparrow$} & \textbf{macro-R$\uparrow$} & \textbf{macro-F1$\uparrow$}\\ \hline
\multirow{4}{*}{one-step} & \multicolumn{2}{|c|}{1} & 0.639 & 0.671 & 0.634   \\
& \multicolumn{2}{|c|}{2} & 0.644 & 0.674 & 0.642 \\
& \multicolumn{2}{|c|}{3} & 0.663 & 0.686 & 0.651   \\
& \multicolumn{2}{|c|}{SHRM} & 0.677 & 0.707 & 0.670   \\
\hline
two-step & \multicolumn{2}{|c|}{SHRM} & \textbf{0.727} & \textbf{0.753} & \textbf{0.723}  \\
\hline
\multicolumn{6}{|c|}{\textbf{Evaluation of RHLVs}}\\ \hline
\multicolumn{3}{|c|}{\textbf{Quantification metrics}} & \multicolumn{3}{c|}{\textbf{Classification metrics}} \\
\hline
$H_{min}/H_{max}$ & \textbf{$RHLV_{avg}$} & \textbf{$RHLV_{segs}$}  & \textbf{macro-P$\uparrow$} & \textbf{macro-R$\uparrow$} & \textbf{macro-F1$\uparrow$}\\ \hline
\checkmark & $\times$ & $\times$ & 0.630 & 0.663 & 0.621   \\
$\times$ & \checkmark & $\times$ & 0.663 & 0.681 & 0.641    \\
$\times$ & $\times$ & \checkmark & \textbf{0.727} & \textbf{0.753} & \textbf{0.723}   \\
\checkmark & \checkmark & $\times$ & 0.710 & 0.743 & 0.700   \\
 \checkmark& $\times$ & \checkmark & 0.720 & 0.730 & 0.704   \\
$\times$ &  \checkmark & \checkmark & 0.722 & 0.746 & 0.716   \\
\checkmark & \checkmark & \checkmark & 0.719 & 0.726 & 0.701    \\
\hline
\multicolumn{6}{|c|}{\textbf{Evaluation of Classifiers}}\\ \hline
\multicolumn{3}{|c|}{\multirow{2}{*}{\textbf{Quantification metrics}}} & \multicolumn{3}{c|}{\textbf{Classification metrics}}\\
\cline{4-6}
\multicolumn{3}{|c|}{} & \textbf{macro-P$\uparrow$} & \textbf{macro-R$\uparrow$} & \textbf{macro-F1$\uparrow$}\\ \hline
\multicolumn{3}{|c|}{kNN} & 0.686 & 0.569 & 0.608\\ \hline
\multicolumn{3}{|c|}{LDA} & 0.646 & 0.597 & 0.612\\ \hline
\multicolumn{3}{|c|}{KDA} & 0.699 & 0.621 & 0.650\\ \hline
\multicolumn{3}{|c|}{NB}  & 0.640 & 0.587 & 0.606\\ \hline
\multicolumn{3}{|c|}{SVM} & \textbf{0.727} & \textbf{0.753} & \textbf{0.723} \\ \hline
\multicolumn{6}{l}{* Mask Strategy -1: bounding rectangle that covers the target vertebra as the mask.}\\
\multicolumn{6}{l}{-2: resize the bounding rectangle to 1.2 times its original height.}\\
\multicolumn{6}{l}{-3: mask that covers the continuous three vertebrae centered on the target.}\\
\end{tabular}
}
\end{table}

\subsection{Evaluation Metrics}
We employ four performance metrics to quantitatively evaluate the proposed generative method, including Structural Similarity Index Measure (SSIM), Peak Signal-to-Noise Ratio (PSNR), Dice score, and Relative Height Difference Ratio (RHDR). RHDR is defined as follows:
\begin{equation}\label{eq10}
    RHDR = \frac{|H_{gen}-H_{ori}|}{H_{gen}}\times 100\%
\end{equation}
SSIM and PSNR are commonly used to evaluate the quality of generated images. Dice score is used in the target vertebrae segmentation task, and RHDR is proposed to evaluate the relative height difference between the predicted and the actual vertebrae. Please note that only normal vertebrae in the dataset are evaluated for generation task. A lower RHDR indicates that the generation can successfully restore the non-fractured vertebral heights. For binary classification task, we use three metrics including sensitivity, accuracy, and F1-score. For multi-classification, we select macro-precision, macro-recall, and macro-F1, ensuring fair evaluation on class-imbalanced datasets.

\subsection{Ablation Study}
\subsubsection{Effectiveness of AHVS, HGAM, EEM, and preprocessing}
\begin{figure}[!t]
\centerline{\includegraphics[width=0.9\columnwidth]{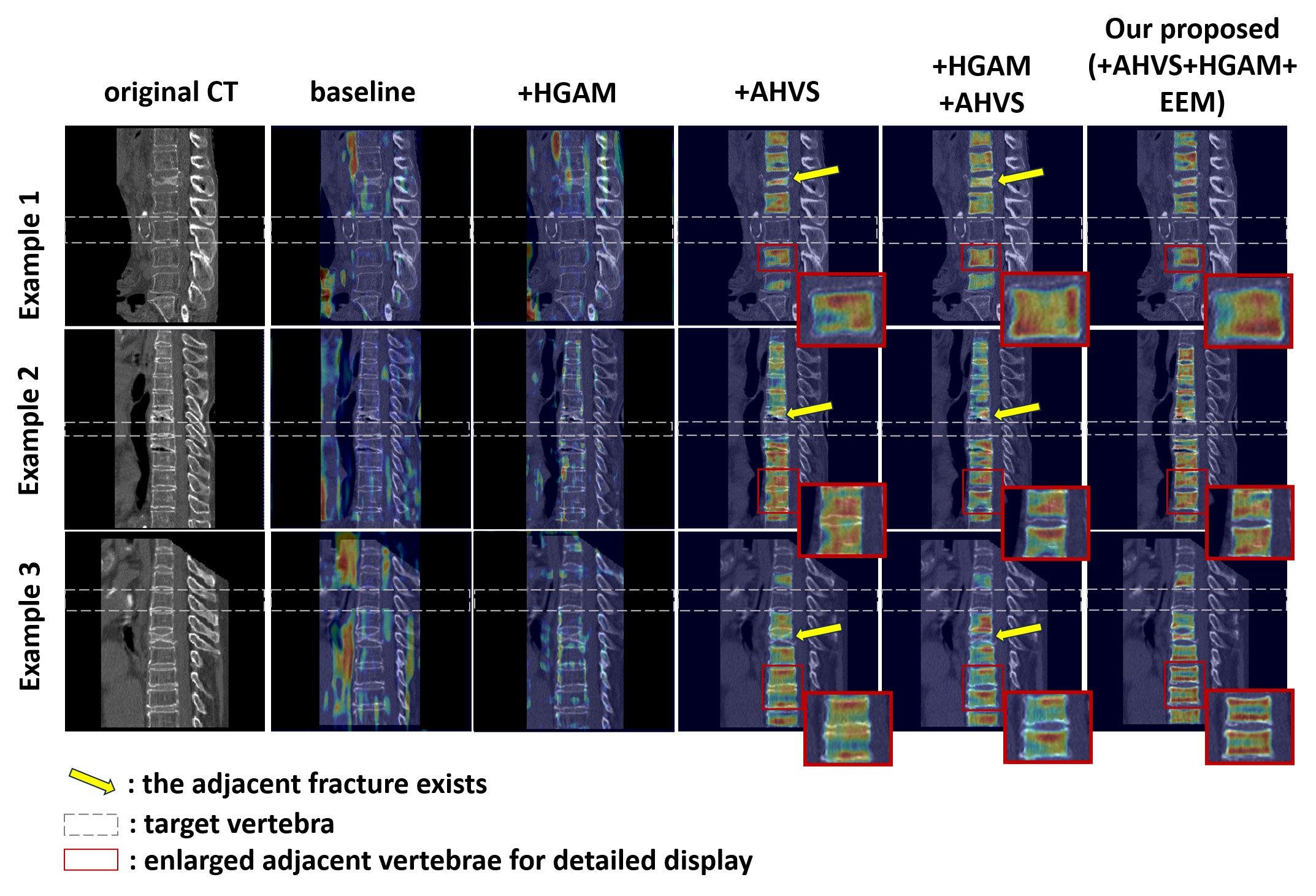}}
\caption{Grad-CAM++ mapping indicates the attention of generation models.}
\label{CAM}
\end{figure}

The ablation results of the proposed modules are shown in \Cref{tab:ablation}. The baseline excludes all three modules. AHVS adds non-fractured vertebrae segmentation Dice loss to the overall loss. HGAM inputs the Grad-CAM++ mapping of into the generator. EEM adds the MSE loss of edge mappings to the overall loss. The quantitative results show that each module improves the baseline in both the generation and classification tasks. It should be noted that macro-F1 improves greatly after adding AHVS, indicating it helps model focus on the adjacent vertebrae to learn useful characteristics. The combination of AHVS and HGAM achieves better results, indicating the effectiveness of healthy vertebrae attention mapping. Additionally, adding EEM significantly boosts RHDR and macro-F1 by improving the morphological details. We visualize the attention map of the generators using Grad-CAM++, as demonstrated in \Cref{CAM}. Adding AHVS helps the model focus on the adjacent vertebrae. Adding HGAM further reduces the influence of the adjacent fractures, as indicated by yellow arrows. EEM helps the generation focus more on the edge of the adjacent vertebrae, as demonstrated in red box. The experiments results of preprocessing in \Cref{tab:ablation} demonstrate that both straightening and de-pedicle are essential for the generation improvement and accurate classification.

\begin{figure*}[!t]
\begin{center}
\includegraphics[width=1.8\columnwidth]{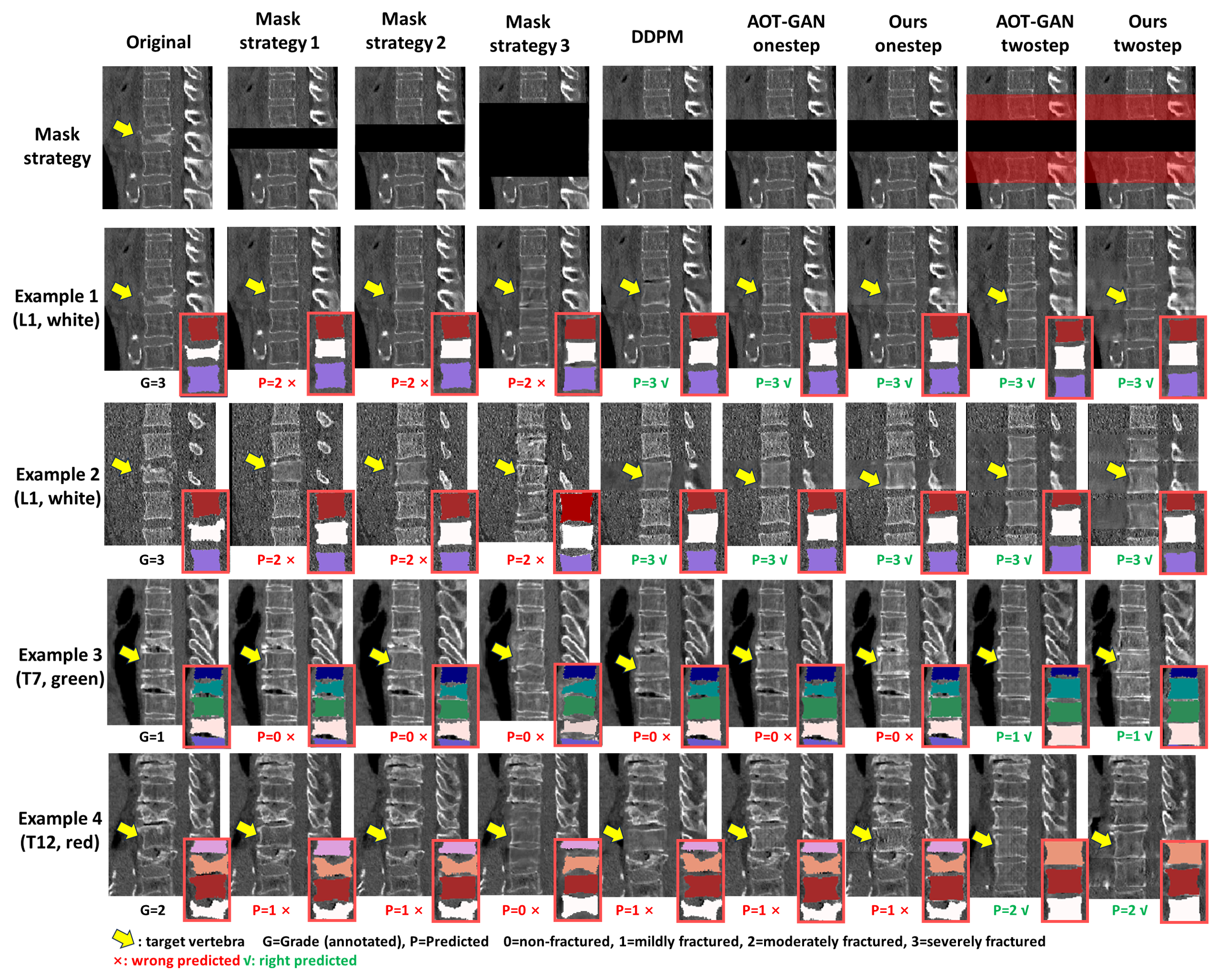}
\end{center}
\caption{Generation results using different mask strategies and different models. Note that in the twostep synthesis, the adjacent vertebrae in red bounding boxes are firstly inpainted, and then target one in the black box is inpainted}
\label{mask}
\end{figure*}

\subsubsection{Effectiveness of two-stage synthesis}
The effectiveness of the coarse-to-fine two-stage synthesis network is evaluated in \Cref{tab:ablation}. Results show that the two-stage architecture outperforms the one-stage approach in both generation and classification tasks. The one-stage method, which combines generation and segmentation in a single step, produces lower-quality images and less accurate classifications due to task competition. In contrast, the two-stage synthesis progressively refines the generated images, leading to higher quality outputs and improved classification performance. Additionally, the sagittal view demonstrates superior performance compared to the coronal view and fused results, highlighting its importance in vertebral fracture analysis.

\subsubsection{Effectiveness of SHRM and iterative synthesis}
We compare different mask strategies in vertebra inpainting, as illustrated in \Cref{tab:evaluation}. The results show that SHRM outperformed other mask strategies and two-step synthesis obtained great improvement than one-step synthesis. The qualitative results are illustrated in \Cref{mask}. Although strategy 1 and 2 can produce realistic vertebral images, the height is limited to the mask size. For strategy 3, we get more space for generation. However, if there are severely compressed segments like Example 3 and 4 in \Cref{mask}, the model outputs poor generation quality, and the vertebral heights are not restored accurately. Additionally, these three mask strategies (1,2,3) depends on the original vertebrae size, bringing the height cues of into the generation. In the proposed SHRM (ours onestep and twostep), we use the fixed-height mask, which provides enough space for generation and also eliminate the original height cues. As shown in Example 1 and 2, both ours one-step and ours two-step methods can restore the severely compressed vertebrae to reasonable shapes and heights. However, for Example 3 and 4, ours one-step output is affected by the adjacent fractures, leading to flawed pseudo-healthy vertebrae. Ours two-step method first inpaints the adjacent vertebrae of the target, alleviating the influence the adjacent fractures.

\subsubsection{Ablation for discriminators}
Results in \Cref{tab:ablation} indicate that incorporating global and local CT discriminators enhances synthesis quality and classification performance. Adding a segmentation discriminator further improves Dice scores and RHDR, facilitating more accurate fracture classification.

\begin{table*}[t]
\centering
\caption{Evaluation of different models on OVCFs grading}
\label{tab:evaluation_all}
\scalebox{0.85}{
\begin{tabular}{|c|c|c|c|c|c|c|c|c|}
\hline
\multicolumn{9}{|c|}{\textbf{Verse 2019 dataset}}\\ \hline
\multicolumn{2}{|c|}{\multirow{2}{*}{\textbf{Models}}} & \multicolumn{3}{c|}{\textbf{Binary-classification}} & \multicolumn{3}{c|}{\textbf{Multi-classification}} & \multirow{2}{*}{\textbf{Parameters}} \\
\cline{3-8}
\multicolumn{2}{|c|}{} & \textbf{Accuracy$\uparrow$} & \textbf{Sensitivity$\uparrow$} & \textbf{F1-score$\uparrow$} & \textbf{macro-P$\uparrow$} & \textbf{macro-R$\uparrow$} & \textbf{macro-F1$\uparrow$} & \\ 
\hline
Regression-based& MLP3 & 0.435±0.043& 0.765±0.035& 0.369±0.046&   0.446±0.034    & 0.399±0.031       & 0.419±0.037        & 0.01 MB \\ \hline
\multirow{11}{*}{Classification-based} & VGG19& 0.725±0.051& 0.654±0.035& 0.623±0.044&   0.639±0.020    & 0.574±0.022       & 0.592±0.017       & 19.11 MB            \\ 
& EfficientNet B2& 0.740±0.029& 0.681±0.045& 0.671±0.032&  0.691±0.051     & 0.564±0.031      & 0.606±0.022       &  7.35 MB            \\ 
& ResNet50       & 0.730±0.024& 0.661±0.041& 0.648±0.037&  0.692±0.025        & 0.570±0.016         & 0.601±0.020         & 22.43 MB               \\ 
& DenseNet161      & 0.721±0.030& 0.667±0.031& 0.641±0.013&   0.665±0.017      & 0.603±0.015         & 0.622±0.016         & 25.26 MB              \\ 
& Base-ViT-16\cite{dosovitskiy2020image}       & 0.736±0.034& 0.634±0.041& 0.662±0.030&    0.622±0.018      & 0.539±0.041         & 0.568±0.030        & 82.11 MB               \\ 
& Base-ViT-32\cite{dosovitskiy2020image}       & 0.734±0.027& 0.615±0.029& 0.629±0.031&    0.551±0.034      & 0.503±0.038         & 0.509±0.022    &  83.48 MB               \\
& Large-ViT-16\cite{dosovitskiy2020image}       & 0.729±0.041& 0.613±0.035& 0.635±0.024&    0.573±0.047      & 0.501±0.033        & 0.516±0.024         & 291.46 MB              \\
& Large-ViT-32\cite{dosovitskiy2020image}      & 0.740±0.022& 0.620±0.040& 0.641±0.023&     0.639±0.031    & 0.534±0.011         & 0.565±0.010         & 289.63 MB               \\
& 3D ResNet      & 0.748±0.032 & 0.601±0.045& 0.643±0.032&     0.627±0.055     & 0.557±0.027        & 0.580±0.034         & 44.05 MB               \\ 
& 2D SupCon-SENet\cite{wei2022faint}     & 0.765±0.040& 0.645±0.035& 0.689±0.011&    0.701±0.018    & 0.634±0.026       & 0.651±0.021       & 25.09 MB          \\ 
& 3D SupCon-SENet\cite{wei2022faint} & 0.780±0.031& 0.679±0.045& 0.710±0.014&  \underline{0.720±0.026}    & 0.636±0.034       & 0.667±0.028        & 46.73 MB         \\ \hline
\multirow{3}{*}{Generation-based} & DDPM\cite{wang2024feasibilitystudydiffusionbasedmodel} & 0.810±0.000 & 0.742±0.000 & 0.732±0.000 &   0.701±0.000    & 0.675±0.000       & 0.675±0.000  &  527.2 MB\\ 
& AOT-GAN\cite{yan2021agg} & \underline{0.823±0.000} & \underline{0.771±0.000} & \underline{0.770±0.000} &  0.710±0.000    & \underline{0.707±0.000}       &\underline{0.692±0.000}  &  32.38 MB\\ 
& HealthiVert-GAN & \textbf{0.834±0.000} & \textbf{0.800±0.000}& \textbf{0.780±0.000}& \textbf{0.727±0.000} & \textbf{0.753±0.000} & \textbf{0.723±0.000}  &9.24 MB\\ \hline
\multicolumn{9}{|c|}{\textbf{In-house dataset}}\\ \hline
\multicolumn{2}{|c|}{\multirow{2}{*}{\textbf{Models}}} & \multicolumn{3}{c|}{\textbf{Binary-classification}} & \multicolumn{3}{c|}{\textbf{Multi-classification}} & \multirow{2}{*}{\textbf{Parameters}} \\
\cline{3-8}
\multicolumn{2}{|c|}{} & \textbf{Accuracy$\uparrow$} & \textbf{Sensitivity$\uparrow$} & \textbf{F1-score$\uparrow$} & \textbf{macro-P$\uparrow$} & \textbf{macro-R$\uparrow$} & \textbf{macro-F1$\uparrow$} & \\
\hline
Regression-based& MLP3 & 0.440±0.049& 0.730±0.045& 0.406±0.050&   0.455±0.036    & 0.415±0.037      & 0.430±0.039        & 0.01 MB \\ \hline
\multirow{11}{*}{Classification-based} & VGG19& 0.715±0.059 & 0.638±0.027 & 0.644±0.050 & 0.612±0.019 & 0.580±0.023 & 0.568±0.020 & 19.11 MB \\
& EfficientNet B2 & 0.710±0.033 & 0.654±0.038 & 0.660±0.038 & 0.622±0.056 & 0.582±0.029 & 0.571±0.020 & 7.35 MB \\
& ResNet50 & 0.740±0.024 & 0.668±0.022 & 0.680±0.037 & 0.637±0.028 & 0.585±0.017 & 0.590±0.022 & 22.43 MB \\
& DenseNet161 & 0.736±0.026 & 0.670±0.034 & 0.675±0.014 & 0.616±0.021 & 0.610±0.017 & 0.620±0.018 & 25.26 MB \\
& Base-ViT-16 \cite{dosovitskiy2020image} & 0.748±0.036 & 0.656±0.047 & 0.660±0.034 & 0.606±0.019 & 0.564±0.049 & 0.588±0.033 & 82.11 MB \\
& Base-ViT-32 \cite{dosovitskiy2020image} & 0.740±0.030 & 0.650±0.030 & 0.658±0.032 & 0.579±0.036 & 0.540±0.042 & 0.582±0.025 & 83.48 MB \\
& Large-ViT-16 \cite{dosovitskiy2020image} & 0.736±0.039 & 0.646±0.038 & 0.668±0.029 & 0.602±0.042 & 0.566±0.038 & 0.594±0.025 & 291.46 MB \\
& Large-ViT-32 \cite{dosovitskiy2020image} & 0.733±0.023 & 0.648±0.041 & 0.675±0.023 & 0.640±0.037 & 0.564±0.012 & 0.601±0.011 & 289.63 MB \\
& 3D ResNet & 0.754±0.037 & 0.625±0.043 & 0.683±0.032 & 0.645±0.047 & 0.578±0.032 & 0.610±0.036 & 44.05 MB \\
& 2D SupCon-SENet \cite{wei2022faint} & 0.780±0.037 & 0.670±0.039 & 0.728±0.012 & 0.670±0.020 & 0.668±0.028 & 0.646±0.024 & 25.09 MB \\
& 3D SupCon-SENet \cite{wei2022faint} & 0.802±0.036 & 0.715±0.051 & 0.748±0.015 & 0.674±0.028 & 0.667±0.033 & 0.674±0.033 & 46.73 MB \\ \hline
\multirow{3}{*}{Generation-based} & DDPM\cite{wang2024feasibilitystudydiffusionbasedmodel} & 0.840±0.000 & 0.762±0.000 & 0.755±0.000 & 0.650±0.000 & 0.701±0.000 & 0.688±0.000 & 527.2 MB\\ 
& AOT-GAN\cite{yan2021agg} & \underline{0.865±0.000} & \underline{0.782±0.000} & \underline{0.780±0.000} & \underline{0.678±0.000} & \underline{0.754±0.000} & \underline{0.701±0.000}   &  32.38 MB\\ 
& HealthiVert-GAN & \textbf{0.872±0.000} & \textbf{0.849±0.000} & \textbf{0.811±0.000} & \textbf{0.680±0.000} & \textbf{0.784±0.000} & \textbf{0.721±0.000} &9.24 MB \\ \hline
\multicolumn{8}{l}{Note: \textbf{Bold} values indicate the best performance for each metric, and \underline{underlined} values indicate the second best.}
\end{tabular}
}
\end{table*}

\begin{table*}[t]
\centering
\caption{Comparison of generation-based models in the in-house dataset}
\label{tab:comparision_generation}
\scalebox{0.85}{
\begin{tabular}{|c|c|c|c|c|c|c|c|c|}
\hline
\multicolumn{2}{|c|}{\multirow{2}{*}{\textbf{Models}}} & \multicolumn{4}{c|}{\textbf{Generation metrics}} & \multicolumn{3}{c|}{\textbf{Multi-classification}} \\
\cline{3-9}
\multicolumn{2}{|c|}{} &\textbf{PSNR$\uparrow$} & \textbf{SSIM$\uparrow$} & \textbf{Dice$\uparrow$} & \textbf{RHDR$\downarrow$} & \textbf{macro-P$\uparrow$} & \textbf{macro-R$\uparrow$} & \textbf{macro-F1$\uparrow$} \\ 
\hline
\multicolumn{2}{|c|}{DDPM\cite{wang2024feasibilitystudydiffusionbasedmodel}} & \textbf{28.36} & \underline{0.924} & 0.876 & 8.22\% & 0.650±0.000 & 0.701±0.000 & 0.688±0.000 \\ 
\hline
\multirow{8}{*}{AOT-GAN\cite{yan2021agg}}& onestep baseline & 27.45& 0.895 & 0.869 & 10.22\% &  0.623±0.000    & 0.659±0.000       & 0.630±0.000 \\ 
& onestep+S& 27.81 & 0.900 & 0.880 & 7.42\% &  0.641±0.000    & 0.697±0.000       & 0.649±0.000    \\ 
& onestep+S+H& 27.93 & 0.915 & 0.887 & 7.10\% &  0.646±0.000    & 0.715±0.000       & 0.657±0.000    \\ 
& onestep+S+H+E & \underline{28.22} & 0.921 & 0.890 & 6.48\% &  0.655±0.000    & 0.724±0.000       & 0.678±0.000      \\ 
\cline{2-9}
& twostep baseline& 27.40 & 0.892 & 0.870 & 9.58\% &   0.632±0.000    & 0.664±0.000       & 0.640±0.000  \\ 
& twostep+S& 27.76 & 0.896 & 0.885 & 6.68\% &   0.660±0.000    & 0.683±0.000       & 0.669±0.000   \\ 
& twostep+S+H& 27.87 & 0.917 & 0.898 & 6.44\% &   0.667±0.000    & 0.691±0.000       & 0.678±0.000     \\ 
& twostep+S+H+E& 28.10 & 0.922 & \underline{0.901} & \underline{6.08\%} &   \underline{0.678±0.000} & \underline{0.754±0.000} & \underline{0.701±0.000}   \\ 
\hline
\multirow{8}{*}{HealthiVert-GAN}& onestep baseline & 27.32& 0.896 & 0.871 & 10.17\% &   0.619±0.000    & 0.661±0.000        & 0.628±0.000  \\ 
& onestep+S & 27.51 & 0.905 & 0.883 & 6.83\% &   0.627±0.000    & 0.694±0.000        & 0.648±0.000    \\ 
& onestep+S+H & 27.72 & 0.916 & 0.891 & 6.67\% &   0.643±0.000    & 0.721±0.000       & 0.661±0.000    \\ 
& onestep+S+H+E & 28.12 & \textbf{0.927} & 0.894 & 6.24\% &   0.645±0.000    & 0.737±0.000       & 0.680±0.000   \\ 
\cline{2-9}
& twostep baseline& 27.45& 0.880 & 0.875 & 9.39\% &   0.626±0.000     & 0.670±0.000        & 0.637±0.000    \\ 
& twostep+S & 27.71 & 0.894 & 0.892 & 6.68\% &   0.650±0.000     & 0.734±0.000        & 0.675±0.000     \\ 
& twostep+S+H & 27.69 & 0.915 & \underline{0.901} & 6.22\% &   0.667±0.000     & \underline{0.754±0.000}        & 0.692±0.000     \\ 
& twostep+S+H+E& 28.03 & 0.923 & \textbf{0.905} & \textbf{5.60\%} & \textbf{0.680±0.000} & \textbf{0.784±0.000} & \textbf{0.721±0.000}  \\ \hline
\multicolumn{9}{l}{+S: Self-adaptive Height Restoration Module; +H: HealthiVert-Guided Attention Mudule; +E: Edge Enhancing Module} \\
\multicolumn{9}{l}{Note: \textbf{Bold} values indicate the best performance for each metric, and \underline{underlined} values indicate the second best.}
\end{tabular}
}
\end{table*}

\subsection{Evaluation Of RHLVs and Different Classifiers}
We compare different inputs to the classifier and evaluate their performance. (1) $H_{min}/H_{max}$: the ratio of minimum to maximum average heights across three segments ($ H_{anterior},H_{middle},H_{posterior} $) of the fractured vertebra. This ratio is independent of generated results; (2) $ RHLV_{avg} $: the average height loss across the entire volumetric vertebra; and (3)$RHLV_{segs}$: the combination of $ RHLV_{anterior}$, $ RHLV_{middle}$, and $ RHLV_{posterior}$. Experiments show that the using only $RHLV_{segs}$ obtains the best classification performance. We also apply different classifiers. Among various classifiers, linear SVM outperforms others, including K Nearest Neighbor (KNN), Linear Discriminant Analysis (LDA), Quadratic Discriminant Analysis (QDA), and Naive Bayes, as demonstrated in \Cref{tab:evaluation}. 

\subsection{Comparison With Other Deep Learning Methods}
\subsubsection{Comparison Methods and Tasks}
We comprehensively compared three model types: (1) Regression-based: a three-layer multilayerperceptron (MLP3) predicting target vertebra height from adjacent vertebrae height array; (2) Classification-based: CNN architectures (VGG19, EfficientNet B2, ResNet50, DenseNet161), Vision Transformer variants (Base/Large-ViT-16/32)\cite{dosovitskiy2020image}, and 3D ResNet (input: [64×64×64] crops with segmentations); (3) Generation-based: AOT-GAN\cite{yan2021agg} and Denoising Diffusion Probabilistic Model (DDPM)\cite{9880056}. Diffusion-based models have been validated in the medical image generation tasks\cite{wang2024diffusion,kazerouni2023diffusion}. Task-specific SupCon-SENet\cite{wei2022faint} (2D/3D versions) was also evaluated according to the original research paper.

For AOT-GAN adaptation, we integrated our modules (SHRM: height prediction head + mask strategy; HGAM: Grad-CAM++ heatmaps; EEM: edge loss) while maintaining HealthiVert-GAN's training protocol. All models were tested on two vertebrae-level tasks: multi-class fracture severity grading (Genant system) and healthy vs. fractured binary classification.

\subsubsection{Evaluation of different models on OVCFs grading}
The experimental results on the Verse2019 and in-house dataset are shown in \Cref{tab:evaluation_all}. For binary classification, all methods except the regression-based model achieved high performance. Among classification-based models, 3D methods generally outperformed 2D methods but required significantly more parameters, leading to higher computational costs. Generation-based models showed superior performance, with HealthiVert-GAN achieving the highest accuracy, sensitivity, and F1-score.

For multi-class classification, HealthiVert-GAN outperformed all other models across all metrics, demonstrating strong ability to distinguish between multiple fractures. ViT models performed poorly, likely due to limited fracture samples. SupCon-SENet, incorporating SupCon loss, outperformed other classification-based models. Compared to classification-based methods, HealthiVert-GAN and AOT-GAN achieved the best and second-best performance with fewer parameters. DDPM also achieved good results but had the largest parameters and longest inference time, limiting its practicality.

\subsubsection{Comparison of generation-based models}
Results comparing three generation-based models are shown in \Cref{tab:comparision_generation}. DDPM achieved the best PSNR and second-best SSIM but performed poorly in anatomical metrics (Dice and RHDR), leading to inferior classification performance. %Its long inference time (5628.0 seconds per scan) also hindered practical clinical application. 
Among GAN-based models including HealthiVert-GAN and AOT-GAN, proposed methods outperformed the baselines, with the modules (SHRM, HGAM, EEM) significantly enhancing both generation and classification performance. The proposed HealthiVert-GAN with two-step generation achieved the best overall performance in generation and classification tasks, with an average inference time of 1.1 seconds, indicating strong clinical applicability.

\iffalse
\begin{table}[ht]
\centering
\caption{Comparison with junior surgeon: in-house dataset}
\label{tab:comparison_junior}
\scalebox{0.80}{
\begin{tabular}{|c|c|c|c|c|c|c|}
\hline
\multirow{2}{*}{\parbox{1.5cm}{\textbf{Method}}} & \multicolumn{4}{c|}{\textbf{Multi-classification}} & \multirow{2}{*}{$\kappa$ value}  & \multirow{2}{*}{\textbf{P value}} \\
\cline{2-5}
& \textbf{Accuracy$\uparrow$} & \textbf{macro-P$\uparrow$} & \textbf{macro-R$\uparrow$} & \textbf{macro-F1$\uparrow$} &&\\ \hline
\parbox{1.5cm}{Junior\\surgeon} & 0.7199 & 0.5035 & 0.5360 & 0.5075 & 0.2987 & 0.000***   \\
\parbox{1.5cm}{HealthiVert\\-GAN} & 0.8414 & 0.6804 & 0.7844 & 0.7210 & 0.5854 &  \\
\hline
\multicolumn{6}{l}{*** McNemar's \(\chi^2\) test.}
\end{tabular}
}
\end{table}

\subsection{Comparison with Junior Surgeon}
We compared the performance of the proposed method with that of a junior surgeon on the multi-classification task. As shown in \Cref{tab:comparison_junior}, HealthiVert-GAN significantly outperformed a junior surgeon in multi-class OVCF grading across all metrics (accuracy: 0.841 vs. 0.720; macro-precision: 0.6804 vs. 0.5035; macro-recall: 0.7844 vs. 0.5360; macro-F1: 0.721 vs. 0.508), demonstrating superior consistency with senior surgeon annotations (Cohen'$\kappa$=0.585 vs. junior $\kappa$=0.299). McNemar's test confirmed the statistical significance of HealthiVert-GAN's superior performance, yielding a p-value $<0.0001$. Confusion matrices (\textbf{Figure S1} (a-b)) reveal reduced misclassification rates achieved by HealthiVert-GAN. 
\fi

\section{DISCUSSION}
% 可以从多个角度分析，主要是审稿人关心的问题，包括了临床诊断和分级的流程、连续骨折的分析、缺少骨折前影像如何模拟的分析、failure cases的分析等等

\begin{figure}[!t]
\centerline{\includegraphics[width=\columnwidth]{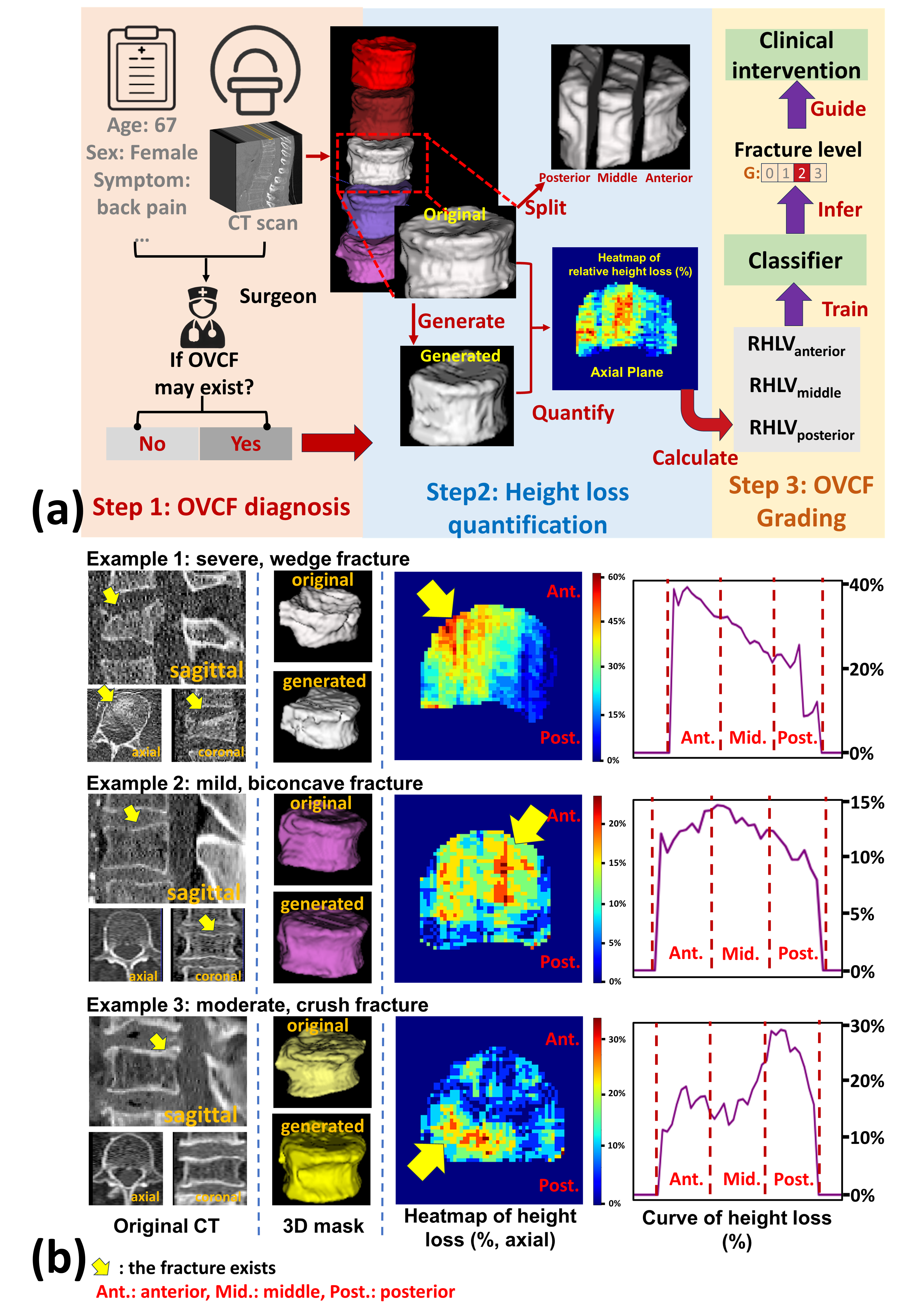}}
\caption{(a) The clinical workflow of our method in OVCF grading. (b) The visualization heatmap of vertebral height loss distribution in axial view, and the curve of height loss. They help surgeons to evaluate the fracture severity across different vertebral regions and determine the fracture morphological types. }
\label{fig:clinical_workflow_distribution}
\end{figure}

\Cref{fig:clinical_workflow_distribution} (a) illustrates the clinical workflow of the proposed method. Accurate quantification and grading are critical steps in determining OVCF patient management, as it directly influences treatment decisions. In clinical practice, moderate or severe fractures require surgical intervention. However, due to the absence of a reliable reference height for the target vertebra, junior surgeons often struggle to estimate the pre-fracture vertebral height, leading to misclassification and potential errors in treatment planning. Our proposed framework addresses this challenge by synthesizing a pseudo-healthy vertebral image that simulates the pre-fracture state. This approach enables precise quantification of vertebral height loss across three anatomical segments (anterior, middle, and posterior), providing a sensitive and interpretable measure of fracture severity. By training an SVM classifier on thousands of vertebrae and corresponding labels, we establish a grading standard that aligns with senior surgeons' annotations. Our method achieves higher inter-rater consistency with senior surgeons compared to junior surgeons (\textbf{Table S1}), demonstrating its potential to help reduce diagnostic variability and support clinical decision-making.

A critical limitation of existing DL-based OVCF grading lies in the lack of 3D interpretability. While methods like Zakharov et al.'s height ratios \cite{zakharov2023interpretable} offer single-slice insights, they fail to capture complex compression patterns (e.g., severely/uniformly vertebral compression or multi-sectional wedge/biconcave fractures). In this study, we introduce a novel paradigm for OVCF quantification and grading by leveraging 3D morphological comparisons between pre-fracture and post-fracture states. As shown in \Cref{fig:clinical_workflow_distribution} (b), our method generates heatmaps of relative height loss in the axial view and plots height loss curves across the vertebra. These visualizations provide a comprehensive assessment of fracture severity and morphology, enabling surgeons to quickly identify regions of maximum height loss and classify fracture types (e.g., wedge, biconcave, or crush fractures). For instance, Example 1 demonstrates a wedge fracture with significant anterior height loss, while Example 2 shows a biconcave fracture with central height loss, and Example 3 indicates a crush fracture affecting the posterior region. Compared to single-plane metrics or traditional fracture grading, our height loss distribution heatmaps and curves offer a panoramic view of vertebral compression across multiple regions and layers.

One of the most concerns in validating the generation is the absence of true pre-fracture CT scans. In clinical practice, patients typically undergo imaging only after experiencing discomfort, making it impossible to obtain true pre-fracture data. To address this limitation, we implemented a three-tier validation protocol: (1) Masked restoration of the healthy vertebrae achieves precise anatomical reconstruction (Dice=0.894, RHDR=6.58\%; \Cref{tab:ablation}), which uses the actual images of themselves as non-fracture references; (2) Vertebral height curve analysis shows a high degree of similarity between actual and restored healthy vertebrae (\textbf{Figure S3} (a)), preserving physiological variation of vertebral heights; (3) Restored fractured vertebrae exhibited height patterns closely aligned with healthy reference curves (\textbf{Figure S3} (a)). These supports the clinical applicability of our generative framework.

\iffalse
\begin{table}[ht]
\centering
\caption{The results of continuous fractures: in Verse 2019 dataset}
\label{tab:continuous_fractures}
\begin{tabular}{|c|c|c|c|c|}
\hline
\multirow{2}{*}{\textbf{Patients}} & \multicolumn{4}{c|}{\textbf{Multi-classification}} \\
\cline{2-5}
& \textbf{Accuracy$\uparrow$} & \textbf{macro-P$\uparrow$} & \textbf{macro-R$\uparrow$} & \textbf{macro-F1$\uparrow$}\\ \hline
All & 0.7812 & 0.7270 & 0.7530 & 0.7230   \\
$\geq3$ & 0.7068 & 0.7880 & 0.7540 & 0.7620   \\
$\geq5$ & 0.5556 & 0.6250 & 0.5210 & 0.5250  \\
\hline
\multicolumn{5}{l}{All: all vertebrae in validation set are included.}\\
\multicolumn{5}{l}{$\geq3$: continuous fractures $\geq3$.}\\
\multicolumn{5}{l}{$\geq5$: continuous fractures $\geq5$.}
\end{tabular}
\end{table}
\fi

Another challenge in vertebral fracture analysis is the impact of adjacent fractures on the accuracy of vertebral generation, particularly in cases involving multiple continuous fractures. Traditional one-step generation methods often fail to accurately restore the target vertebra when adjacent vertebrae are severely compressed. To address this issue, we propose a two-step iterative synthesis strategy. \textbf{Table S2} provides a detailed analysis of model performance across different groups of continuous fractures. While the baseline group (All) achieves an accuracy of 0.7812 and a macro-F1 of 0.7230, the group with $\geq3$ continuous fractures shows a slight decrease in accuracy (0.7068) but maintains or even improves other metrics (e.g., macro-F1: 0.7620). However, performance significantly declines when continuous fractures reach $\geq5$, with macro-F1 decreasing from 0.7230 to 0.5250. Confusion matrices for the groups are illustrated in \textbf{Figure S1} (c-d).

%\begin{figure}[ht]
%\centerline{\includegraphics[width=\columnwidth]{failure_cases_continuous_fractures.png}}
%\caption{(a) Cases of continuous fractures and the different generation results. (b) Failure case of preprocessing (segmentation and straightening). (c) Failure cases of wrong OVCF classification, we choose the T8 and T10 in sub-verse080.}
%\label{fig:failure_cases_continuous_fractures}
%\end{figure}

%这部分可以放到补充材料里面
To evaluate the model's reliability in handling severe continuous fractures, we analyzed two extreme cases from the Verse 2019 dataset: Case 1 with 6 continuous fractures (T12-L5) and Case 2 with 9 continuous fractures (T6-L3), as shown in \textbf{Figure S2} (a). Three generation strategies were compared: one-step, two-step, and continuous generation (restoring other vertebrae sequentially with the target vertebra generated last). The vertebral height curves generated by each strategy revealed that continuous generation most effectively simulated the pre-fracture height curves, even in these extreme cases. The two-step strategy closely matched the continuous generation results, demonstrating its practical applicability. In contrast, the one-step strategy, while still functional, performed less effectively than the other two methods. These findings underscore the effectiveness and importance of iterative synthesis in addressing the challenges posed by adjacent continuously fractures. 

The proposed method, while effective for most challenging cases, has several limitations. First, its performance is dependent on accurate preprocessing, particularly vertebral segmentation and spine straightening. \textbf{Figure S2} (b) shows a failure case where T9 and T10 were misidentified, causing consecutive straightening errors. Although segmentation accuracy was high across most cases (only 5 errors in 497 in-house dataset cases), manual correction is still required to ensure reliable downstream processing. Future work will focus on advanced segmentation techniques to reduce such preprocessing failures.
Second, the model struggles with ambiguous fractures, especially when height loss is at the boundary between normal and mild categories. \textbf{Figure S1} (b) indicates that most classification errors occur between grades 0 and 1. \textbf{Figure S2} (c) provides two examples: a false negative in T8 due to adjacent fracture influence and a false positive in T10 where height loss is ambiguously classified. These findings are consistent with previous studies \cite{zakharov2023interpretable, husseini2020grading}, which reported lower performance in G0 vs G1G2G3 tasks and suggested excluding G1 from evaluations due to high inter-rater uncertainty. The confusion matrix in \textbf{Figure S1} (a) further highlights this uncertainty between junior and senior surgeons. Addressing this requires refining the grading system, possibly through revised reference standards or additional imaging features, with further clinical validation needed to assess the impact on model performance and clinical applicability. 
Third, the model is currently validated on the public Verse2019 dataset and our single-center in-house dataset. To address the limitations of single-center validation, we plan to construct and evaluate our model on a multi-center vertebral CT dataset to enhance its generalizability across diverse populations.

\section{CONCLUSION}
In this study, we introduced HealthiVert-GAN, a novel framework developed for the synthesis of three-dimensional pseudo-healthy vertebral images aimed at enhancing the assessment of OVCFs. By simulating the pre-fracture state of vertebrae, our approach offers a new paradigm in the automated grading and precise quantification of vertebral fractures. To effectively learn the non-fractured characteristics, we propose three auxilary modules: HGAM, SHRM, and EEM, to enhance the authenticity and applicability of the generated images. Additionally, we demonstrated that our proposed framework and modules can be transferred to other GAN-based models to improve baseline performance. We introduced the RHLV metric as a quantifiable standard for assessing vertebral height loss, which, combined with SVM classification, achieved state-of-the-art performance in fracture grading across two datasets. Using the generated three-dimensional pseudo-healthy vertebrae, we can analyze the distribution of height loss across multiple cross-sections and directions. This analysis provides critical information for assessing the impact of vertebral fractures on vertebral stability and aids in determining surgical indications.

%Future work will focus on expanding the clinical applicability of our framework. Current OVCF grading primarily relies on height loss measurements, but additional imaging features such as CT density and texture information may provide complementary insights into fracture risk and stability. Furthermore, to address the limitations of single-center validation, we plan to construct and evaluate our model on a multi-center vertebral CT dataset to enhance its generalizability across diverse populations. 

%\section*{Appendix and the Use of Supplemental Files}
%\footnote{}

%\section*{Acknowledgment}
%Thanks to the support of National Medical Products Administration (NMPA) Key Laboratory for Medical Electrical Equipment.
%\section{References}

%\begin{thebibliography}{00}
\section{References}

\bibliographystyle{ieeetr}
\bibliography{main}

%\end{thebibliography}

\end{document}